\documentclass[aps,PRB,reprint,superscriptaddress,showpacs,preprintnumbers]{revtex4-1}
\usepackage[T1]{fontenc}
\usepackage{times}
\usepackage{graphicx,color}
\usepackage{amsfonts,amsmath,amssymb,amsbsy}
\usepackage[colorlinks=true, linkcolor=blue, citecolor=blue, urlcolor=blue]{hyperref}

\let\mathbf=\boldsymbol

\def\blue#1{\textcolor{blue}{#1}}
\def\black#1{\textcolor{black}{#1}}
\def\emph#1{\textcolor[rgb]{1,0,1}{#1}}
\begin{document}

\title{Static and dynamic properties of bimerons in a frustrated ferromagnetic monolayer}

\author{\href{https://orcid.org/0000-0001-9656-9696}{\black{Xichao Zhang}}}
\thanks{These authors contributed equally to this work.}
\affiliation{School of Science and Engineering, The Chinese University of Hong Kong, Shenzhen, Guangdong 518172, China}

\author{\href{https://orcid.org/0000-0002-1009-3074}{\black{Jing Xia}}}
\thanks{These authors contributed equally to this work.}
\affiliation{School of Science and Engineering, The Chinese University of Hong Kong, Shenzhen, Guangdong 518172, China}

\author{\href{https://orcid.org/0000-0003-2604-8339}{\black{Laichuan Shen}}}
\affiliation{School of Science and Engineering, The Chinese University of Hong Kong, Shenzhen, Guangdong 518172, China}

\author{\href{https://orcid.org/0000-0002-3629-5643}{\black{Motohiko Ezawa}}}
\email[]{ezawa@ap.t.u-tokyo.ac.jp}
\affiliation{Department of Applied Physics, The University of Tokyo, 7-3-1 Hongo, Tokyo 113-8656, Japan}

\author{\href{https://orcid.org/0000-0001-7283-6884}{\black{Oleg A. Tretiakov}}}
\affiliation{School of Physics, The University of New South Wales, Sydney 2052, Australia}

\author{\href{https://orcid.org/0000-0002-3327-9445}{\black{Guoping Zhao}}}
\affiliation{College of Physics and Electronic Engineering, Sichuan Normal University, Chengdu 610068, China}

\author{\href{https://orcid.org/0000-0001-5917-5495}{\black{Xiaoxi Liu}}}
\affiliation{Department of Electrical and Computer Engineering, Shinshu University, 4-17-1 Wakasato, Nagano 380-8553, Japan}

\author{\href{https://orcid.org/0000-0001-5641-9191}{\black{Yan Zhou}}}
\email[]{zhouyan@cuhk.edu.cn}
\affiliation{School of Science and Engineering, The Chinese University of Hong Kong, Shenzhen, Guangdong 518172, China}

\begin{abstract}
Magnetic bimeron is a topological counterpart of skyrmions in in-plane magnets, which can be used as a spintronic information carrier. We report the static properties of bimerons with different topological structures in a frustrated ferromagnetic monolayer, where the bimeron structure is characterized by the vorticity $Q_{\text{v}}$ and helicity $\eta$. It is found that the bimeron energy increases with $Q_{\text{v}}$, and the energy of an isolated bimeron with $Q_{\text{v}}=\pm 1$ depends on $\eta$. We also report the dynamics of frustrated bimerons driven by the spin-orbit torques, which depend on the strength of the dampinglike and fieldlike torques. We find that the isolated bimeron with $Q_{\text{v}}=\pm 1$ can be driven into linear or elliptical motion when the spin polarization is perpendicular to the easy axis. We numerically reveal the damping dependence of the bimeron Hall angle driven by the dampinglike torque. Besides, the isolated bimeron with $Q_{\text{v}}=\pm 1$ can be driven into rotation by the dampinglike torque when the spin polarization is parallel to the easy axis. The rotation frequency is proportional to the driving current density. In addition, we numerically demonstrate the possibility of creating a bimeron state with a higher or lower topological charge by the current-driven collision and merging of bimeron states with different $Q_{\text{v}}$. Our results could be useful for understanding the bimeron physics in frustrated magnets.
\end{abstract}

\date{April 27, 2020}

\preprint{\textit{Phys. Rev. B} \textbf{101}, 144435 (2020). DOI:~\href{https://doi.org/10.1103/PhysRevB.101.144435}{10.1103/PhysRevB.101.144435}}

\maketitle

\section{Introduction}
\label{se:Introduction}

The emerging fields of chiral magnetism and topological spintronics are developed through the study of topologically nontrivial spin textures, which can be found in a variety of magnetic materials~\cite{Nagaosa_NNANO2013, Wiesendanger_NATREVMAT2016, Finocchio_JPD2016, Kang_PIEEE2016, Kanazawa_AM2017, Wanjun_PHYSREP2017, Fert_NATREVMAT2017, Zhou_NSR2018, ES_JAP2018, Zhang_JPCM2020}, and can be used for information encoding in data storage and processing devices~\cite{Nagaosa_NNANO2013, Wiesendanger_NATREVMAT2016, Finocchio_JPD2016, Kang_PIEEE2016, Kanazawa_AM2017, Wanjun_PHYSREP2017, Fert_NATREVMAT2017, Zhou_NSR2018, ES_JAP2018, Zhang_JPCM2020}.
An appealing and well-studied topological spin texture is the magnetic skyrmion in perpendicularly magnetized system, the existence of which was theoretically predicted in 1989~\cite{Bogdanov_1989, Roszler_NATURE2006}, and was experimentally observed in 2009~\cite{Muhlbauer_SCIENCE2009}.
Since the first experimental observation, magnetic skyrmions have been extensively studied due to their particlelike nature~\cite{Lin_PRB2013, Reichhardt_2017} and large potential in advanced electronic and spintronic applications~\cite{Yu_NATURE2010, Du_NCOMMS2015, Jiang_SCIENCE2015, Woo_NMATER2016, MoreauLuchaire_NNANO2016, Matsumoto_SA2016, Wanjun_NPHYS2017, Litzius_NPHYS2017, Pollard_NCOMMS2017, Hrabec_NC2017, Woo_NC2017, Woo_NatElect2018, Woo_NC2018, Ma_NL2019, Nozaki_APL2019}.
So far, the creation, annihilation, and manipulation of magnetic skyrmions have been realized in magnetic multilayers at room temperature~\cite{Jiang_SCIENCE2015, Woo_NMATER2016, MoreauLuchaire_NNANO2016, Wanjun_NPHYS2017, Litzius_NPHYS2017, Pollard_NCOMMS2017, Hrabec_NC2017, Woo_NC2017, Woo_NatElect2018, Woo_NC2018, Ma_NL2019, Nozaki_APL2019}, and a lot of skyrmion-based device applications have been proposed~\cite{Sampaio_NNANO2013, Tomasello_SREP2014, Xichao_SREP2015B, Senfu_NJP2015, Xichao_NCOMMS2016, Bourianoff_AIP2016, Tomasello_JPD2017, Muller_NJP2017, Yangqi_NANO2017, Lisai_NANO2017, Prychynenko_PRAPPL2018, Zhang_PRB2016} and even demonstrated in room-temperature experiments~\cite{Guoqiang_NL2017, Zazvorka_NN2019, Woo_NE2020}.

Similar to the magnetic skyrmion, the magnetic bimeron is a topological spin texture in in-plane magnetized system~\cite{Zhang_JPCM2020,Xichao_SREP2015B,Lin_PRB2015,Kharkov_PRL2017,Leonov_PRB2017,Fernandes_SSC2019,Gobel_PRB2019,Kim_PRB2019,Moon_PRApplied2019,Ezawa_PRB2011,Chmiel_NM2018,Kolesnikov_SR2018,Murooka_SR2020,Shen_PRL2020,Lu_2020,ES_PRB2020}, which carries an integer topological charge $Q$ and can be regarded as a topological counterpart of skyrmions.
An isolated bimeron with $Q=1$ consists of a pair of merons with $Q=1/2$, however, merons are unstable solutions so that they exist in magnetic systems in the form of pairs~\cite{Lin_PRB2015,Kharkov_PRL2017,Leonov_PRB2017,Chmiel_NM2018,Kolesnikov_SR2018,Fernandes_SSC2019,Gobel_PRB2019,Kim_PRB2019,Moon_PRApplied2019,Murooka_SR2020,Lu_2020,Shen_PRL2020,ES_PRB2020}.
Note that the concept of meron was originated in classical field theory~\cite{DeAlfaro_PLB1976}, and has been studied in different fields, such as quantum Hall systems~\cite{Ezawa_2010,Ezawa_2011}.

In magnetism, bimerons are of great importance to electronic and spintronic applications. For example, they can be used to carry binary digital data and can be driven into motion by spin torques, which are useful for building racetrack-type memory and information computing devices~\cite{Gobel_PRB2019,Kim_PRB2019,Moon_PRApplied2019,Murooka_SR2020,ES_PRB2020}.
Recent reports have suggested the existence and manipulation of bimerons in different magnetic materials, including ferromagnets~\cite{Ezawa_PRB2011,Lin_PRB2015,Leonov_PRB2017,Gobel_PRB2019,Kim_PRB2019,Moon_PRApplied2019,Murooka_SR2020,ES_PRB2020}, antiferromagnets~\cite{Kolesnikov_SR2018,Fernandes_SSC2019,Shen_PRL2020}, and frustrated magnets~\cite{Kharkov_PRL2017}.
In particular, a lattice of bimerons (i.e., a square lattice of merons and antimerons) has been observed in chiral-lattice magnet Co$_{8}$Zn$_{9}$Mn$_{3}$ by using Lorentz transmission electron microscope in 2018~\cite{Yu_Nature2018}.

However, the dynamics of magnetic bimerons driven by different external forces as well as the static properties of different forms of bimerons still remain elusive, especially for the bimerons in frustrated magnetic systems. Some most recent studies have focused on the existence and manipulation of skyrmions in frustrated magnetic systems~\cite{Kharkov_PRL2017,Okubo_PRL2012, Leonov_NCOMMS2015, Batista_Review2016, Lin_PRB2016A, Hayami_PRB2016A, Rozsa_PRL2016, Desplat_PRB2019, Leonov_NCOMMS2017, Xichao_NCOMMS2017, Yuan_PRB2017, Hou_AM2017, Hu_SR2017, Malottki_SR2017, Liang_NJP2018, Diep_Entropy2019, Kurumaji_SCIENCE2019, Xia_PRApplied2019, Zarzuela_PRB2019, Lohani_PRX2019, Diep_2020}. Therefore, it is also imperative to study bimerons in frustrated magnetic systems, of which the physical properties are essential for designing future bimeron-based device applications.

In this work, we report the static properties of bimerons with different values of topological charge $Q$ and helicity $\eta$ in a frustrated ferromagnetic monolayer system, where the bimerons are stabilized by a delicate competition between frustrated exchange interactions, in-plane magnetic anisotropy, and long-range dipole-dipole interaction. We also report the dynamics of frustrated bimerons driven by the spin-orbit torques, where both the dampinglike and fieldlike torques are taken into account.

\section{Methods}
\label{se:Methods}

We consider a frustrated ferromagnetic (FM) monolayer with three competing Heisenberg exchange interactions based on the $J_{1}$-$J_{2}$-$J_{3}$ model on a simple square lattice~\cite{Lin_PRB2016A,Xichao_NCOMMS2017,Xia_PRApplied2019}.
The Hamiltonian is expressed as
\begin{align}
\mathcal{H}=&-J_1\sum_{\substack{<i,j>}}\boldsymbol{m}_i\cdot\boldsymbol{m}_j-J_2\sum_{\substack{\ll i,j\gg}}\boldsymbol{m}_i\cdot\boldsymbol{m}_j \notag \\
&-J_3\sum_{\substack{\lll i,j\ggg}}\boldsymbol{m}_i\cdot\boldsymbol{m}_j-K\sum_{\substack{i}}{(m^x_i)^{2}}+H_{\text{DDI}},
\label{eq:Hamiltonian}
\end{align}
where $\boldsymbol{m}_{i}$ represents the normalized spin at the site $i$, $|\boldsymbol{m}_{i}|=1$.
$\left\langle i,j\right\rangle$,
$\left\langle\left\langle i,j\right\rangle\right\rangle$, and
$\left\langle\left\langle\left\langle i,j\right\rangle\right\rangle\right\rangle$
run over all the nearest-neighbor (NN), next-nearest-neighbor (NNN), and next-next-nearest-neighbor (NNNN) sites in the FM monolayer, respectively.
$J_1$, $J_2$, and $J_3$ are the coefficients for the NN, NNN, and NNNN Heisenberg exchange interactions, respectively.
$K$ is the easy-axis magnetic anisotropy constant, and the $x$-axis direction is defined as the easy axis.
$H_{\text{DDI}}$ represents the dipole-dipole interaction (DDI), i.e., the demagnetization term.

The spin dynamics is simulated by using the object oriented micromagnetic framework (OOMMF)~\cite{OOMMF} with periodic boundary condition (PBC) and our extension modules for the $J_{1}$-$J_{2}$-$J_{3}$ classical Heisenberg model~\cite{Lin_PRB2016A,Xichao_NCOMMS2017,Xia_PRApplied2019}, including two modules for the NNN and NNNN exchange interactions.
The OOMMF conjugate gradient minimizer is employed for the spin relaxation simulation, which locates local minima in the energy surface through direct minimization techniques~\cite{OOMMF}.

\begin{figure}[t]
\centerline{\includegraphics[width=0.485\textwidth]{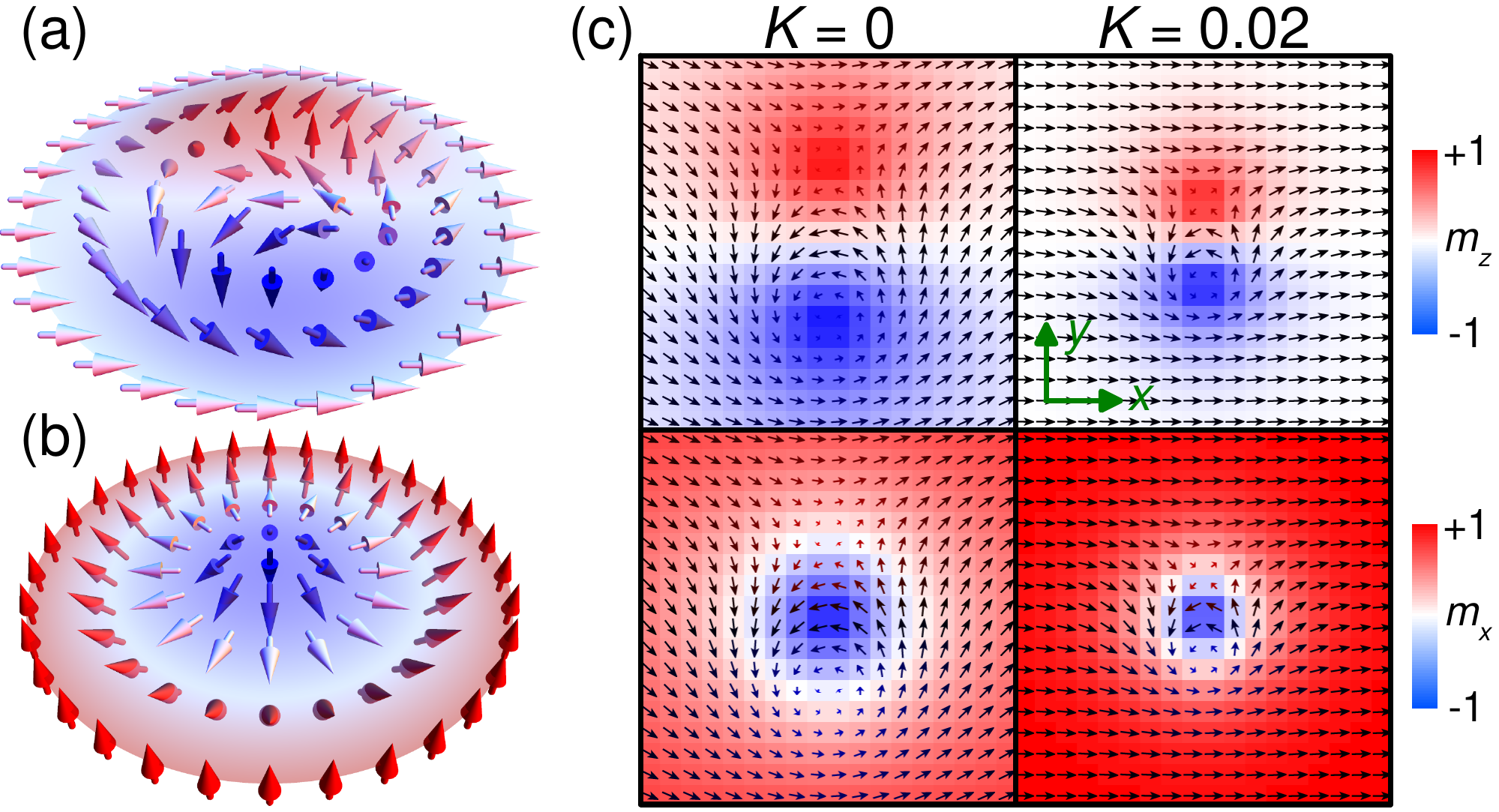}}
\caption{%
(a) Illustration of an isolated magnetic bimeron with $Q_{\text{v}}=+1$ and $\eta=0$.
(b) Illustration of an isolated magnetic skyrmion with $Q_{\text{v}}=+1$ and $\eta=0$.
(c) Zoomed top view of a relaxed isolated bimeron in the FM monolayer in the absence ($K=0$) or presence ($K=0.02$) of in-plane magnetic anisotropy. The relaxed bimeron has $Q_{\text{v}}=+1$ and $\eta=0$.
The arrows represent the magnetization directions. The out-of-plane component of magnetization ($m_z$) is color coded in (a), (b), and the upper panel of (c). The in-plane component of magnetization ($m_x$) is color coded in the lower panel of (c).
}
\label{FIG1}
\end{figure}

The time-dependent spin dynamics is governed by the Landau-Lifshitz-Gilbert (LLG) equation~\cite{OOMMF}
\begin{equation}
\frac{d\boldsymbol{m}}{dt}=-\gamma_{0}\boldsymbol{m}\times\boldsymbol{h}_{\rm{eff}}+\alpha\left(\boldsymbol{m}\times\frac{d\boldsymbol{m}}{dt}\right),
\label{eq:LLG}
\end{equation}
where $\boldsymbol{m}$ represents the normalized spin,
$\boldsymbol{h}_{\rm{eff}}=-\delta\mathcal{H}/\delta\boldsymbol{m}$ is the effective field,
$t$ is the time,
$\alpha$ is the Gilbert damping parameter,
and $\gamma_0$ is the absolute gyromagnetic ratio.
As for the spin-orbit torques, which can be generated by harnessing the spin Hall effect of a heavy metal substrate~\cite{Finocchio_JPD2016,Kang_PIEEE2016,Wanjun_PHYSREP2017,Fert_NATREVMAT2017,ES_JAP2018,Zhang_JPCM2020,Sampaio_NNANO2013,Tomasello_SREP2014,Ado_PRB2017}, we consider here both the dampinglike torque $\tau_{1}$ and fieldlike torque $\tau_{2}$, given as
\begin{equation}
\tau_{1}=-u\boldsymbol{m}\times(\boldsymbol{m}\times\boldsymbol{p}), \; \; \; \tau_{2}=-\xi u(\boldsymbol{m}\times\boldsymbol{p}),
\label{eq:SOT}
\end{equation}
where $u=|\frac{\gamma_{0}\hbar}{\mu_{0}e}|\frac{j \theta_{\text{SH}}}{2a M_{\text{S}}}$ is the spin torque coefficient and $\xi$ is the strength of the fieldlike torque $\tau_{2}$.
$\hbar$ is the reduced Planck constant, $e$ is the electron charge, $\mu_{0}$ is the vacuum permeability constant, $a$ is the thickness of the FM monolayer (here also as the lattice constant),
$j$ is the applied current density, $\theta_{\text{SH}}$ is the spin Hall angle, and $M_{\text{S}}$ is the saturation magnetization.
$\boldsymbol{p}$ denotes the polarization direction of the spin current. 
$\tau_1$ and $\tau_2$ are added to the right-hand side of Eq.~(\ref{eq:LLG}) when the considered spin-orbit torques are nonzero.

In this work, the default values for the NN, NNN, and NNNN exchange interactions are given as~\cite{Xichao_NCOMMS2017,Xia_PRApplied2019}: $J_1=30$ meV, $J_2=-0.8$ (in units of $J_1=1$), and $J_3=-0.6$ (in units of $J_1=1$).
Note that we have carried out metastability diagram simulations showing the parameters dependence and found that the bimeron states can exist in the relaxed sample for a wide range of $J_2$, $J_3$, and $K$ parameters.
The default values for other parameters are~\cite{Xichao_NCOMMS2017,Xia_PRApplied2019}:
$K=0.02$ (in units of $J_{1}/a^{3}=1$),
$\theta_{\text{SH}}=0.2$,
$\alpha=0.3$,
$\gamma_0=2.211\times 10^{5}$ m A$^{-1}$ s$^{-1}$, and
$M_{\text{S}}=580$ kA m$^{-1}$.
The default geometry of the FM monolayer is set as $20 \times 20 \times 0.4$ nm$^3$, and the cubic cells generated by the spatial discretization is of $0.4 \times 0.4 \times 0.4$ nm$^3$, i.e., the lattice constant is $a=0.4$ nm.
Namely, the default monolayer sample consists of $50 \times 50$ spins.

We define the topological charge $Q$ in the continuum limit by the formula~\cite{Nagaosa_NNANO2013,Zhang_JPCM2020}
\begin{equation}
Q={\frac{1}{4\pi}}\int\boldsymbol{m}(\boldsymbol{r})\cdot\left(\partial_{x}\boldsymbol{m}(\boldsymbol{r})\times\partial_{y}\boldsymbol{m}(\boldsymbol{r})\right)d^{2}\boldsymbol{r}.
\label{eq:Q_m}
\end{equation}
The topological charge $Q$ counts how many times $\boldsymbol{m}(\boldsymbol{r})$ wraps $2$-sphere as the coordinate $(x,y)$ spans the whole planar space. We parametrize the spin texture as
\begin{equation}
\boldsymbol{m}(\boldsymbol{r})=\boldsymbol{m}(\theta,\phi)=(\cos\theta,\sin\theta\cos\phi,\sin\theta\sin\phi),
\label{eq:m_theta_phi_bimeron}
\end{equation}
with
\begin{equation}
\phi=Q_{\text{v}}\psi+\eta,
\label{eq:Q_eta}
\end{equation}
where $\psi$ is the azimuthal angle in the $y$-$z$ plane ($0\le\psi<2\pi$) and we assume that $\theta$ rotates $\pi$ for spins from the bimeron center to sample edge. 
Hence, $Q_{\text{v}}=\frac{1}{2\pi}\oint_{C}d \phi$ is the vorticity and $\eta$ is the helicity defined mod $2\pi$.
The internal structure of a bimeron is described by both $Q_{\text{v}}$ and $\eta$, and we may index a bimeron state as $(Q_{\text{v}},\eta)$ in this work. Note that $\eta=0$ is identical to $\eta=2\pi$.

\begin{figure}[t]
\centerline{\includegraphics[width=0.485\textwidth]{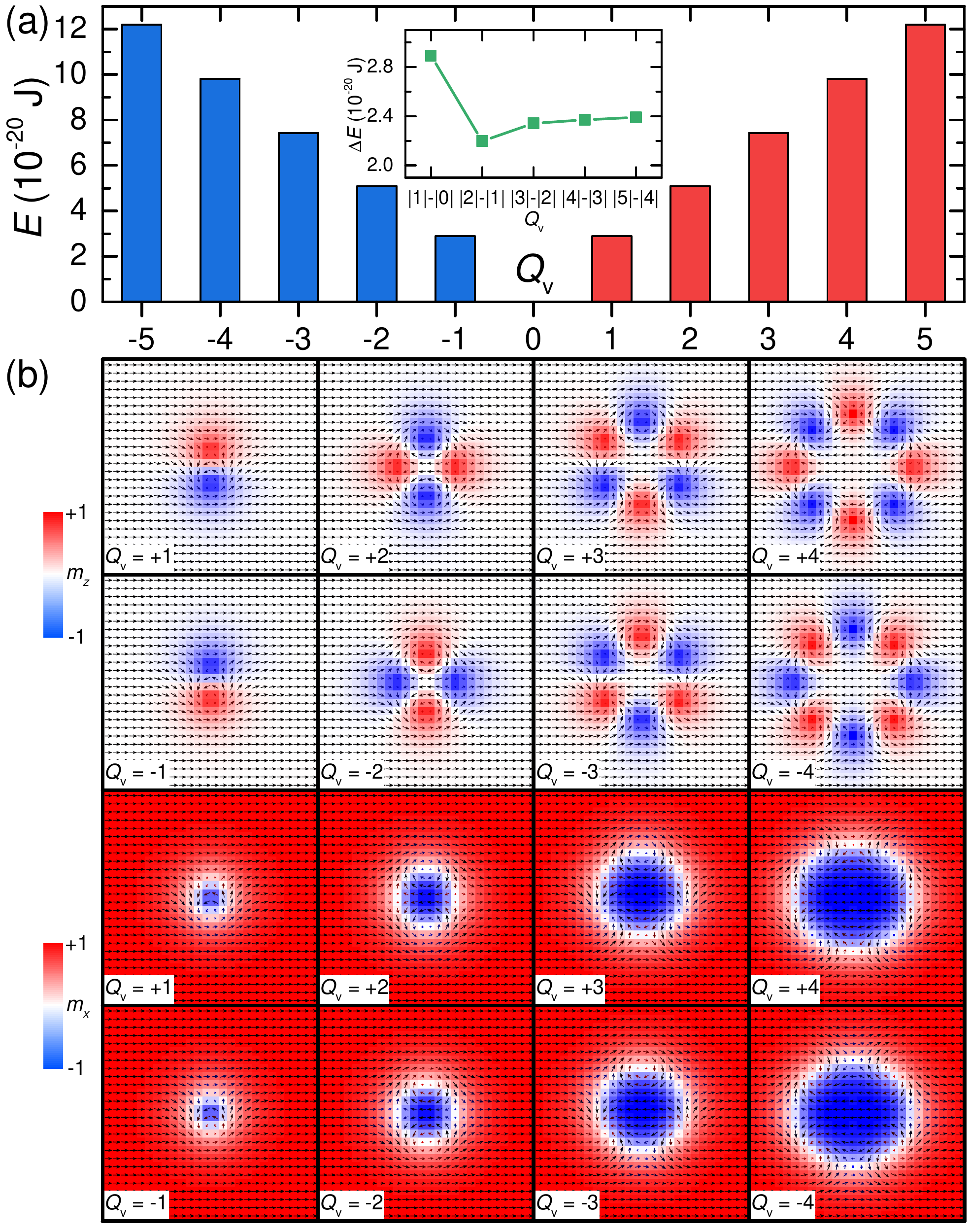}}
\caption{%
(a) Total energy as a function of $Q_{\text{v}}$ for relaxed bimeron states in the FM monolayer with in-plane magnetic anisotropy ($K=0.02$). Inset shows the total energy difference between the bimeron states with consecutive vorticity numbers. Note that $Q_{\text{v}}=0$ corresponds to the FM state, i.e., all magnetization is aligned along the $+x$ direction due to the anisotropy. The total energy of the FM state equals $1.45\times 10^{-23}$ J, which is significantly smaller than that of bimeron states.
(b) Top view of relaxed bimeron states with different $Q_{\text{v}}$ in the FM monolayer with in-plane anisotropy ($K=0.02$).
}
\label{FIG2}
\end{figure}

\begin{figure}[t]
\centerline{\includegraphics[width=0.485\textwidth]{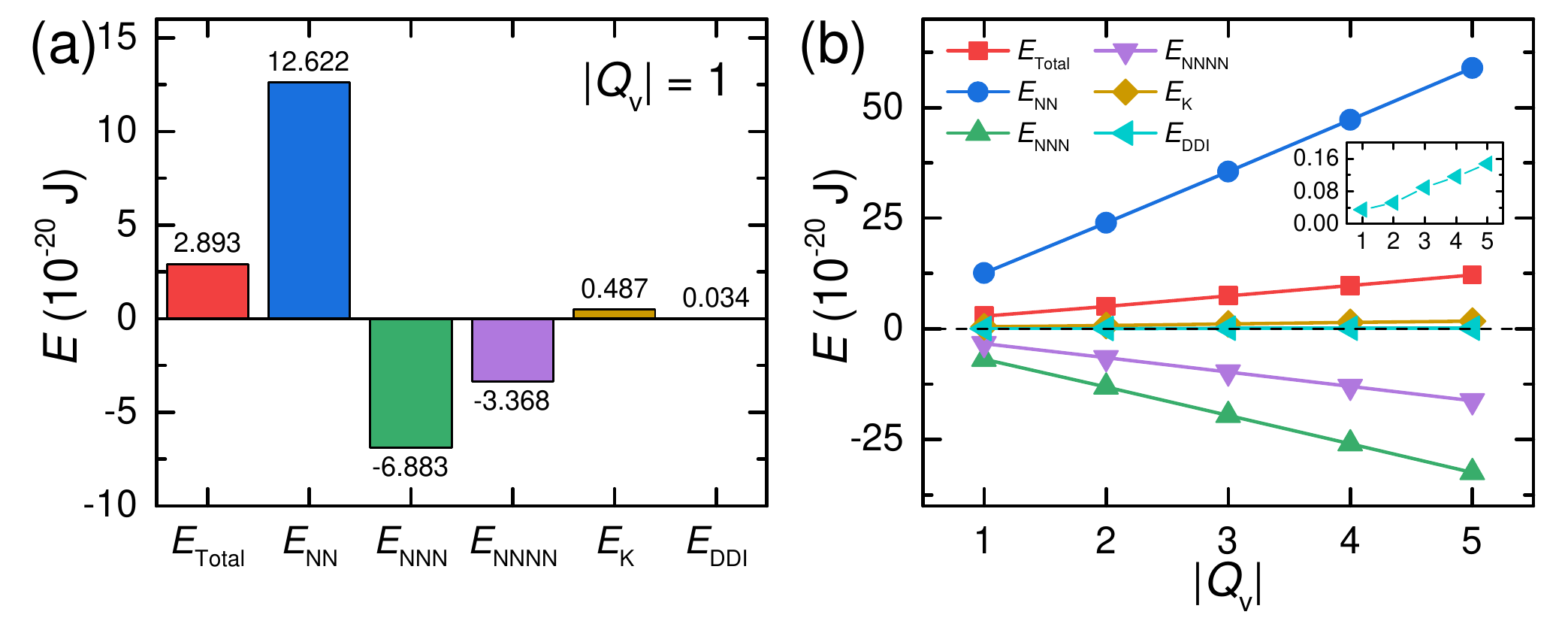}}
\caption{%
(a) The contribution of different energy terms to the total energy $E_{\text{Total}}$ of a relaxed isolated bimeron with $Q_{\text{v}}=\pm 1$, including the NN exchange energy $E_{\text{NN}}$, NNN exchange energy $E_{\text{NNN}}$, NNNN exchange energy $E_{\text{NNNN}}$, anisotropy energy $E_{\text{K}}$, and DDI energy $E_{\text{DDI}}$.
(b) Energies of a relaxed isolated bimeron as functions of $\left|Q_{\text{v}}\right|$. Inset shows the zoomed view of the DDI energy $E_{\text{DDI}}$, the value of which is much smaller than others.
}
\label{FIG3}
\end{figure}

\section{Results and Discussion}
\label{se:Results}

\subsection{Static properties of frustrated bimerons}
\label{se:SPB}

The bimeron [see Fig.~\ref{FIG1}(a)] in in-plane magnetic system is a topological counterpart of skyrmion [see Fig.~\ref{FIG1}(b)] in out-of-plane magnetic system, both of which carry integer topological charge $Q$ and can be stabilized in FM monolayers and ultra-thin films with exchange frustration~\cite{Kharkov_PRL2017,Okubo_PRL2012, Leonov_NCOMMS2015, Batista_Review2016, Lin_PRB2016A, Hayami_PRB2016A, Rozsa_PRL2016, Desplat_PRB2019, Leonov_NCOMMS2017, Xichao_NCOMMS2017, Yuan_PRB2017, Hou_AM2017, Hu_SR2017, Malottki_SR2017, Liang_NJP2018, Diep_Entropy2019, Kurumaji_SCIENCE2019, Xia_PRApplied2019, Zarzuela_PRB2019, Lohani_PRX2019, Diep_2020}.
Note that the skyrmion with $Q=1$ has one out-of-plane core, while the bimeron with $Q=1$ has two opposite out-of-plane cores.
Here we study the static properties of isolated bimerons in a FM monolayer with competing exchange interactions, in-plane magnetic anisotropy, and DDI.

We first simulate the metastability diagram of bimerons for different values of NNN exchange interaction $J_{2}$, NNNN exchange interaction $J_3$, and in-plane easy-axis anisotropy $K$. We consider both randomly distributed initial spin configuration and the initial spin configuration of an isolated bimeron with $(+1,0)$ at the monolayer center. The simulated metastability diagrams are given in Supplemental Material (see Supplementary Figs.~\blue{1}-\blue{14})~\cite{SM}. It is found that metastable isolated bimerons with $Q_{\text{v}}=\pm 1$ and bimeron states with higher topological charges $\left|Q_{\text{v}}\right|>1$ can exist in certain ranges of $J_{2}$ and $J_{3}$, assuming $J_{1}=1$.
Note that the metastability diagrams of frustrated bimerons are similar to that of frustrated skyrmions (see Refs.~\onlinecite{Xichao_NCOMMS2017,Kharkov_PRL2017}).
However, the skyrmion states cannot be spontaneously formed in the absence of the out-of-plane magnetic anisotropy, while the metastable bimeron states can be formed even in the absence of the in-plane magnetic anisotropy.
The reason could be the effect of demagnetization (i.e., the DDI), which is very well approximated by the in-plane magnetic anisotropy in monolayers and ultra-thin films, and favors in-plane spin configurations. Thus, it can contribute to the stability of bimerons of compact shape and small core size (compare Supplementary Fig.~\blue{10} with Supplementary Fig.~\blue{14})~\cite{SM}.
Besides, the size of relaxed metastable bimeron with $Q_{\text{v}}=\pm 1$ decreases with increasing magnitude of the in-plane magnetic anisotropy $K$, as shown in Fig.~\ref{FIG1}(c). In the absence of in-plane anisotropy ($K=0$), the size of the relaxed bimeron with $Q_{\text{v}}=\pm 1$ equals about $8$ times the lattice constant (i.e., $3.2$ nm), which is defined by the diameter of the circle with $m_{x}=0$. However, for the system with $K=0.02$, the bimeron size is about $5$ times the lattice constant (i.e., $2$ nm).
On the other hand, both the upper and lower thresholds of $J_{2}$ and $J_{3}$ for stabilizing bimeron states increase with increasing $K$~\cite{SM}.

We also investigate bimeron states with different $Q_{\text{v}}$ and $\eta$ by relaxing the FM monolayer with $J_{2}=-0.8$, $J_{3}=-0.6$, and $K=0.02$, where a bimeron solution with $(Q_{\text{v}},\eta)$ at the monolayer center is given as the initial spin configuration. After the relaxation of the system, the spin configuration is a stable or metastable solution.
As shown in Fig.~\ref{FIG2}, the total energy of the relaxed bimeron state with $\eta=0$ increases with the absolute value of its vorticity number $Q_{\text{v}}$.
For the bimeron states with the same values of $\left|Q_{\text{v}}\right|$ and $\eta$, their total energies are identical.
For example, the total energies of relaxed isolated bimerons with $(\pm 1,0)$ equal $2.89\times 10^{-20}$ J.
The contribution of different energy terms to the total energy of an isolated bimeron with $Q_{\text{v}}=\pm 1$ is given in Fig.~\ref{FIG3}. It can be seen that in the in-plane magnetic system, the DDI energy contribution is quite small, while the competition between the NN, NNN, and NNNN exchange energy contributions is obvious [see Fig.~\ref{FIG3}(a)]. The NN exchange energy, anisotropy energy, and DDI energy increase with increasing $\left|Q_{\text{v}}\right|$, while the NNN and NNNN exchange energies decrease with increasing $\left|Q_{\text{v}}\right|$ [see Fig.~\ref{FIG3}(b)].

Note that one could refer to the bimeron with $Q_{\text{v}}<0$ as the antibimeron, although the total energy does not depend on the sign of $Q_{\text{v}}$. The difference between the bimeron and antibimeron with the same absolute value of $Q_{\text{v}}$ lies in the out-of-plane spin configuration [see Fig.~\ref{FIG2}(b)]. It is worth mentioning that the difference between a skyrmion and an antiskyrmion lies in the in-plane spin configuration~\cite{Koshibae_NCOMMS2016,Kovalev_Review2018,Ritzmann_NE2018,Potkina_2019}.

We also study the energy of a single isolated bimeron with $(\pm 1, 0)$ for different geometric parameters. As the bimeron state is a localized solution, its total energy is independent of the lateral sample's size, while the total bimeron energy increases with the thickness of the sample (see Supplementary Figs.~\blue{15} and~\blue{16})~\cite{SM}. For the monolayer sample with a fixed side length, the bimeron energy shows asymptotic behavior as the lattice constant $a$ is reduced (see Supplementary Figs.~\blue{17})~\cite{SM}. We note that in our monolayer model, the bimeron energy is identical for samples with odd and even numbers of spins in the in-plane direction (see Supplementary Figs.~\blue{18})~\cite{SM}.

On the other hand, the total energy difference between two bimerons with consecutive numbers of $Q_{\text{v}}$ is given in the inset of Fig.~\ref{FIG2}.
It can be seen that the total energy of an isolated bimeron with $Q_{\text{v}}=\pm 1$ is larger than the energy increase between two bimerons with consecutive numbers of $Q_{\text{v}}$. Therefore, considering the topological conservation and energy minimization of the bimeron system, we predict that a bimeron with $Q_{\text{v}}=1$ can be merged with a bimeron state with $Q_{\text{v}}=n$ ($n$ is a natural number, $n>1$), creating a bimeron state with a higher $Q_{\text{v}}=n+1$. Indeed, the merging of a bimeron with $Q_{\text{v}}= -1$ and a bimeron state with $Q_{\text{v}}= n$ will lead to the formation of a bimeron with lower $Q_{\text{v}}= n-1$.
We will numerically examine this prediction later (see Sec.~\ref{se:DPB}).
Indeed, we point out that the total energy difference grow with $n$ and may eventually become larger than the energy of an isolated bimeron with $Q_{\text{v}}=\pm 1$. This phenomenon is known in nuclear physics when the nuclei with large atomic weight (i.e., large number of protons and neutrons) eventually become unstable~\cite{Krane_1987}. In our case, these protons and neutrons are bimerons with $Q_{\text{v}}=\pm 1$.

\begin{figure}[t]
\centerline{\includegraphics[width=0.485\textwidth]{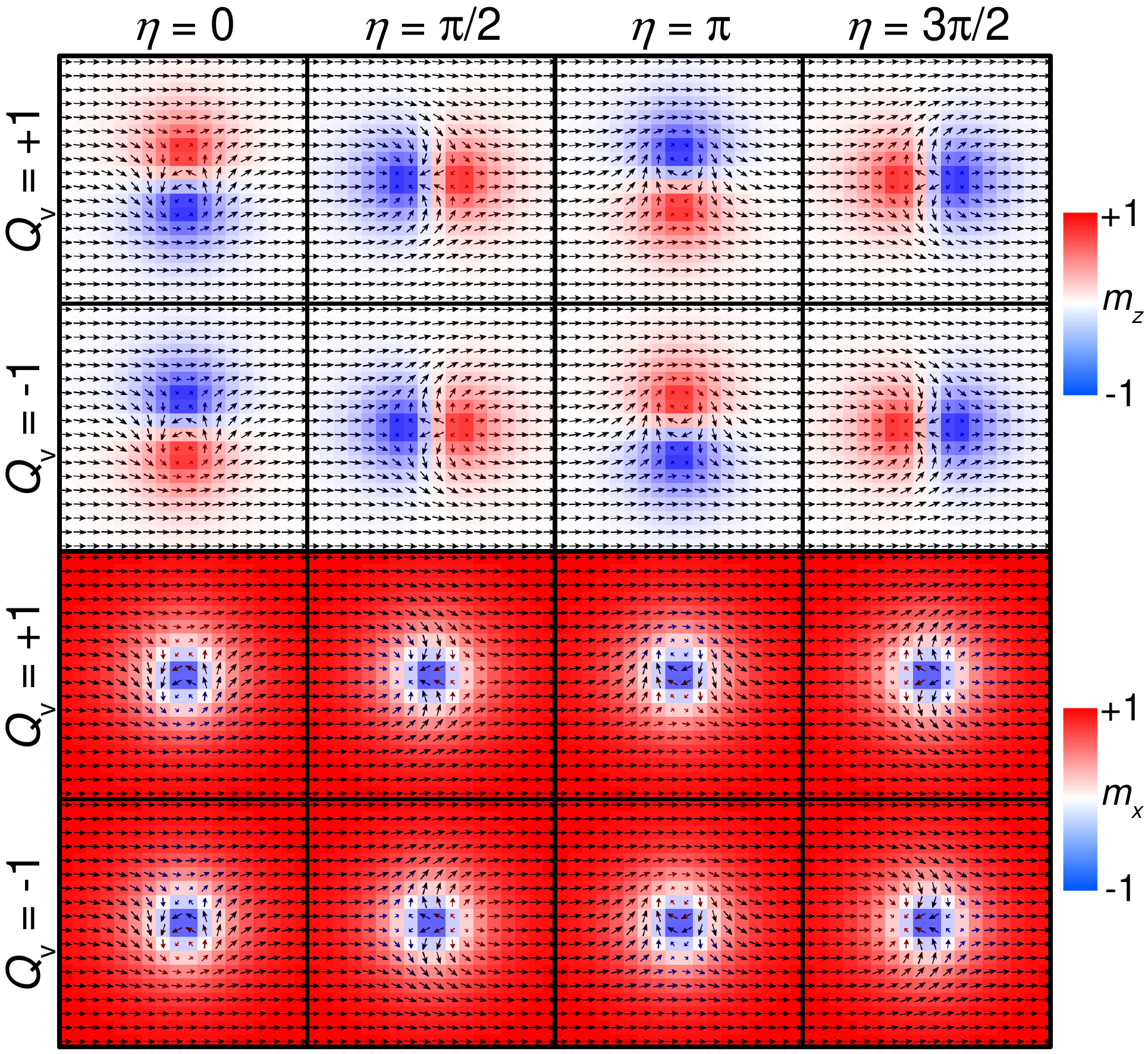}}
\caption{%
Top view of relaxed isolated bimerons with $Q_{\text{v}}=\pm 1$ and different $\eta$ in the FM monolayer with in-plane anisotropy ($K=0.02$).
}
\label{FIG4}
\end{figure}

\begin{figure}[t]
\centerline{\includegraphics[width=0.485\textwidth]{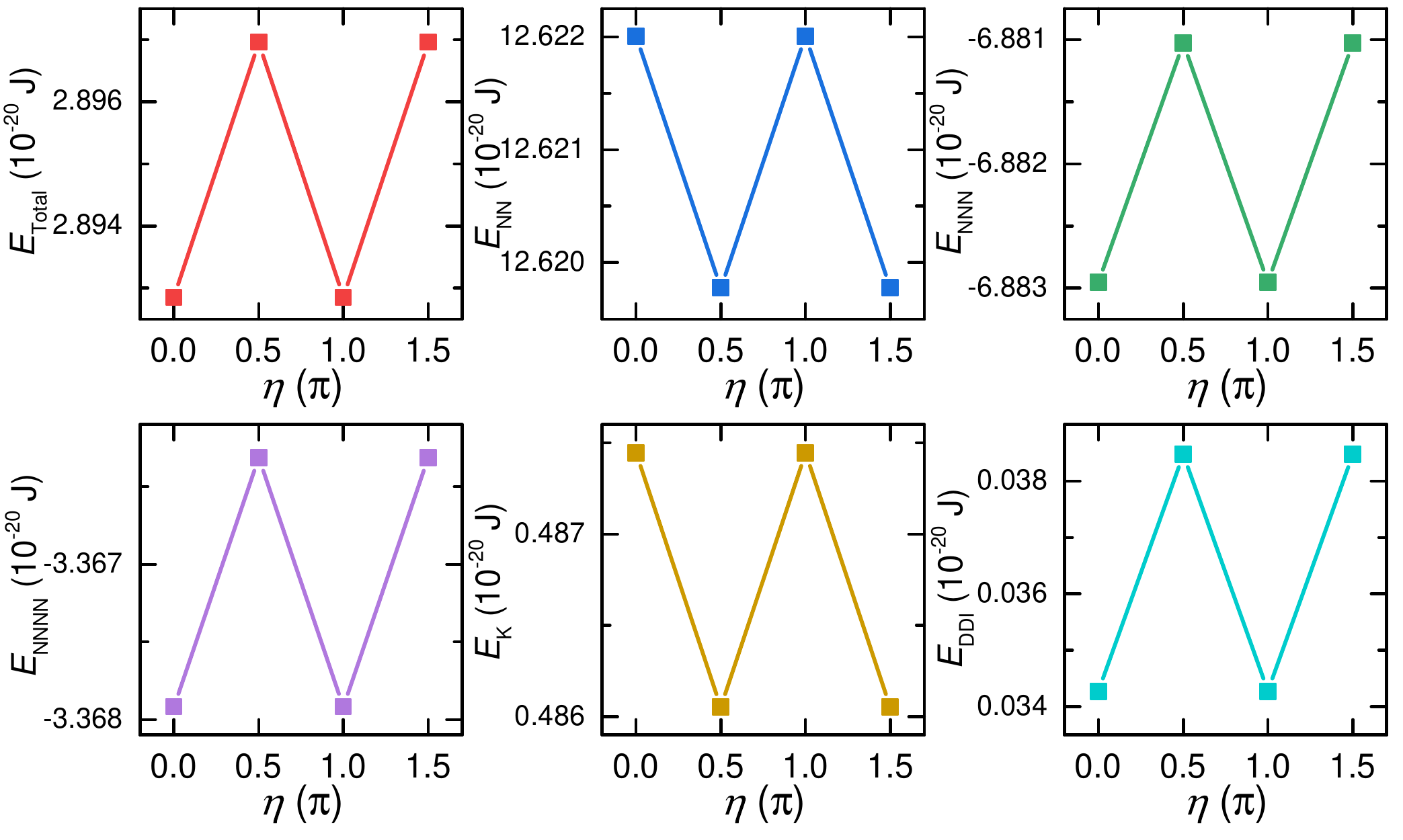}}
\caption{%
Energies of relaxed isolated bimerons with $Q_{\text{v}}=\pm 1$ and different $\eta$ in the FM monolayer with in-plane anisotropy ($K=0.02$). Snapshots of relaxed isolated bimerons are given in Fig.~\ref{FIG4}.
}
\label{FIG5}
\end{figure}

\begin{figure}[t]
\centerline{\includegraphics[width=0.485\textwidth]{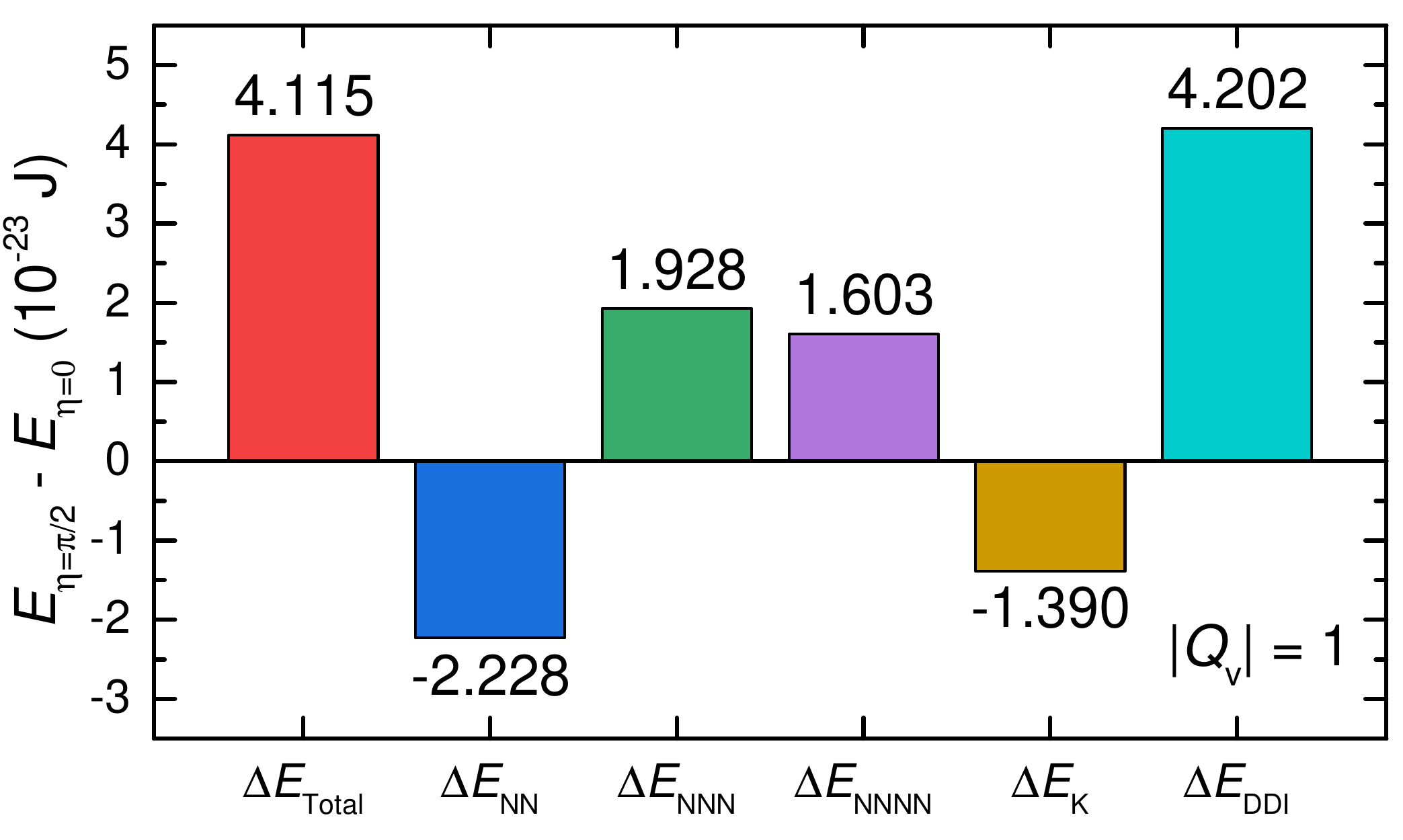}}
\caption{%
Total energy difference between the relaxed isolated bimerons with $(+1,\pi/2)$ and $(+1,0)$.
Note that the energy of the relaxed isolated bimeron with $(\pm 1,\pi/2)$ is identical to that with $(\pm 1,3\pi/2)$, and the energy of the relaxed isolated bimeron with $(\pm 1,0)$ is identical to that with $(\pm 1,\pi)$ (see Fig.~\ref{FIG5}).
}
\label{FIG6}
\end{figure}

\begin{figure}[t]
\centerline{\includegraphics[width=0.450\textwidth]{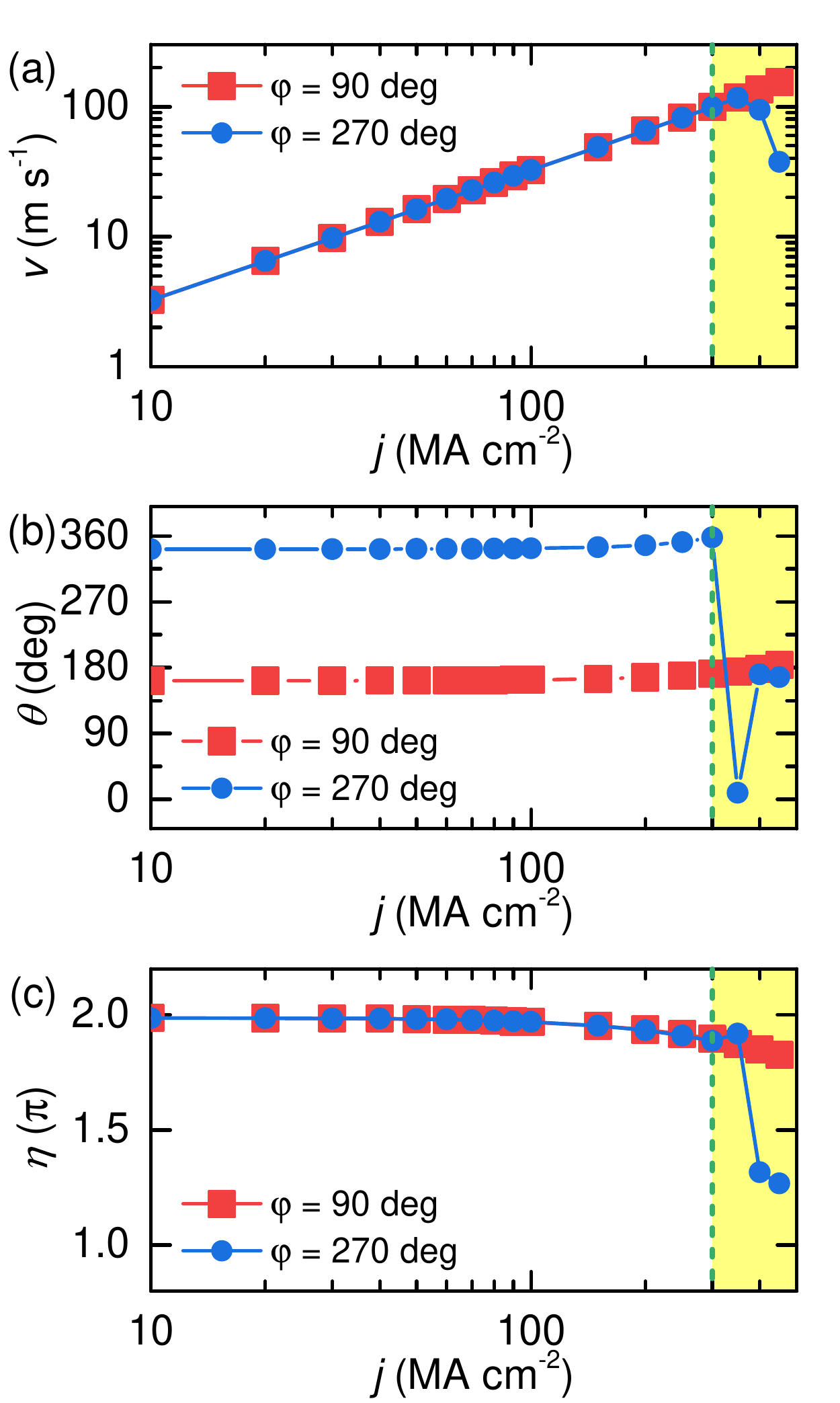}}
\caption{%
(a) Bimeron velocity $v$ as a function of driving current density. The isolated bimeron with $(+1,2\pi)$ is driven by the dampinglike torque $\tau_{1}$. The spin polarization direction is aligned along the $+y$ (i.e., $\varphi=90^{\circ}$) or $-y$ (i.e., $\varphi=270^{\circ}$) direction. Namely, the spin polarization is perpendicular to the easy axis of anisotropy.
(b) Bimeron Hall angle $\theta$ as a function of driving current density. $\theta=\arctan(v_{y}/v_{x})$.
(c) Bimeron helicity $\eta$ as a function of driving current density.
Note that the yellow area indicates that the dynamics of bimeron driven by $\tau_{1}$ with $\varphi=270^{\circ}$ changes from the linear motion with a constant velocity to the elliptical motion with an oscillating velocity. Thus, the bimeron velocity, bimeron Hall angle and bimeron helicity in the yellow area are average values for the bimeron driven by $\tau_{1}$ with $\varphi=270^{\circ}$. The bimeron driven by $\tau_{1}$ with $\varphi=90^{\circ}$ always shows a linear motion for the given range of driving current density.
}
\label{FIG7}
\end{figure}

\begin{figure*}[t]
\centerline{\includegraphics[width=0.800\textwidth]{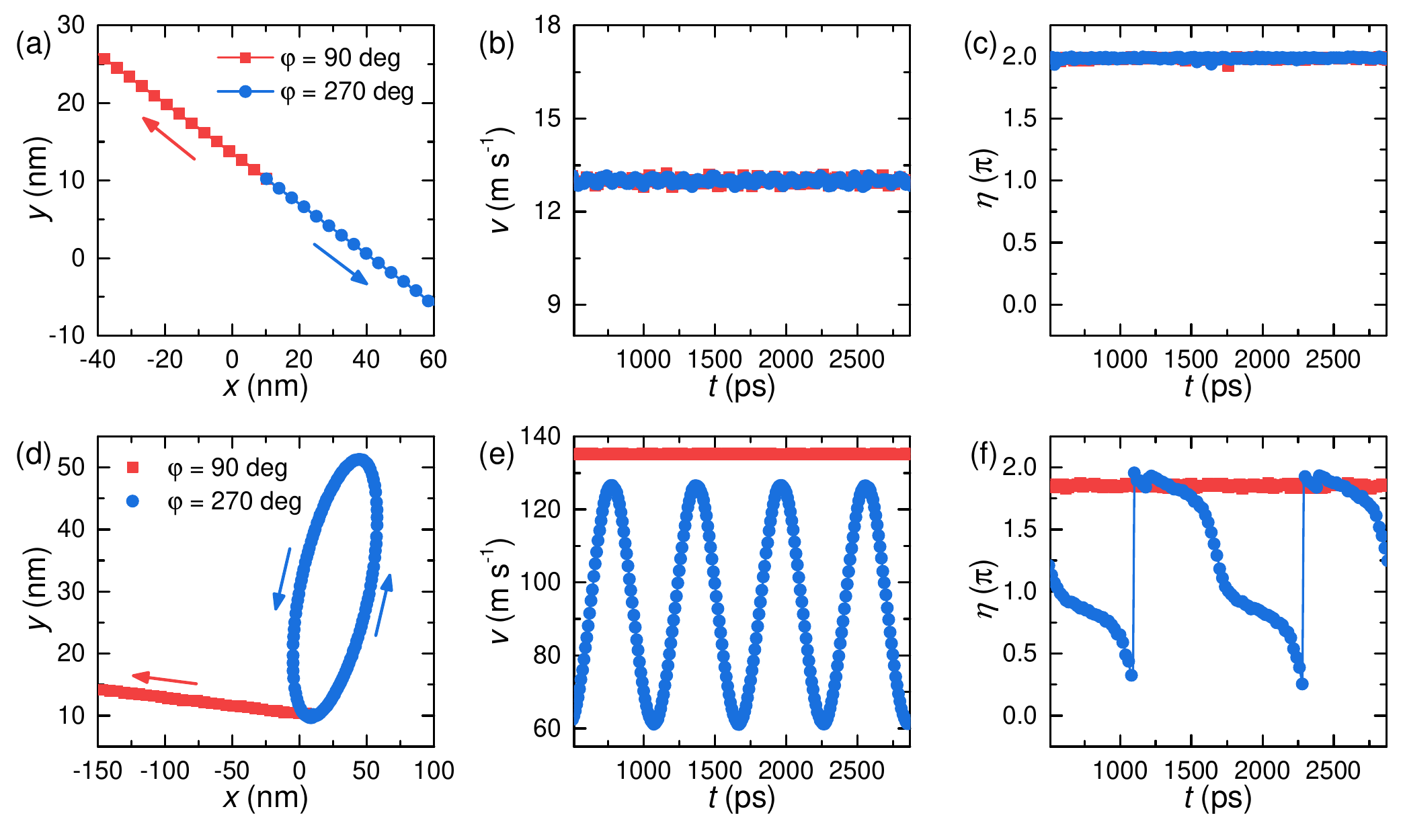}}
\caption{%
(a) Trajectories of the bimeron with $(+1,2\pi)$ driven by the dampinglike torque $\tau_{1}$. The spin polarization direction is aligned along the $+y$ (i.e., $\varphi=90^{\circ}$) or $-y$ (i.e., $\varphi=270^{\circ}$) direction. Namely, the spin polarization is perpendicular to the easy axis of anisotropy. The driving current density $j=40$ MA cm$^{-2}$. The arrow denotes the motion direction.
(b) Bimeron velocity $v$ as a function of time corresponding to (a).
(c) Bimeron helicity $\eta$ as a function of time corresponding to (a).
(d) Trajectories of the bimeron with $(+1,2\pi)$ driven by the dampinglike torque $\tau_{1}$ at $j=400$ MA cm$^{-2}$.
(e) Bimeron velocity $v$ as a function of time corresponding to (d).
(f) Bimeron helicity $\eta$ as a function of time corresponding to (d).
Once the elliptical motion of bimeron is induced, its velocity $v$ and helicity $\eta$ vary with time. 
}
\label{FIG8}
\end{figure*}

\begin{figure}[t]
\centerline{\includegraphics[width=0.485\textwidth]{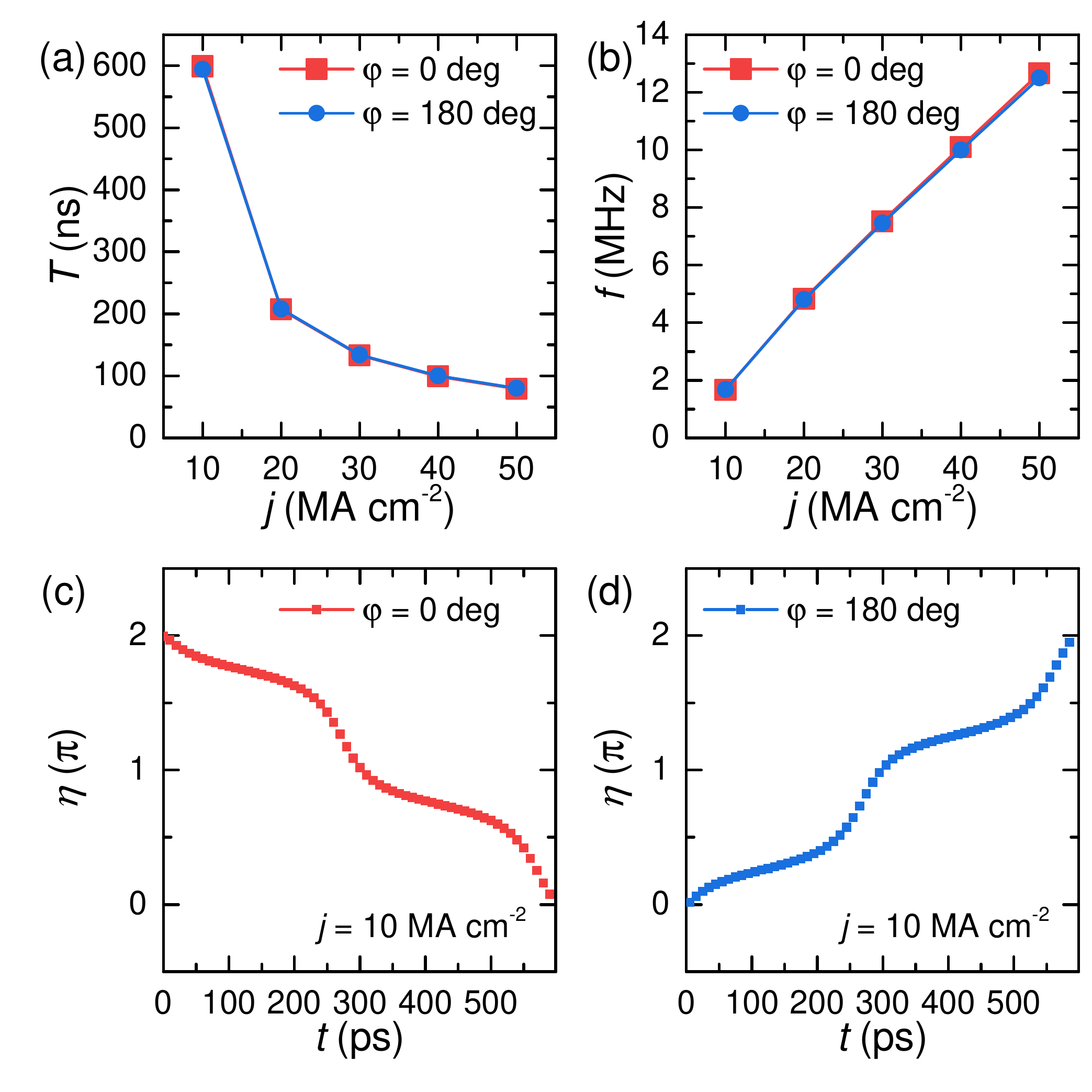}}
\caption{%
(a) Period of bimeron rotation $T$ as a function of driving current density. The bimeron with $Q_{\text{v}}=+1$ is driven by the dampinglike torque $\tau_{1}$. The spin polarization direction is aligned along the $+x$ (i.e., $\varphi=0^{\circ}$) or $-x$ (i.e., $\varphi=180^{\circ}$) direction. Namely, the spin polarization is parallel to the easy axis of anisotropy.
(b) Frequency of bimeron rotation $f$ as a function of driving current density.
(c) Bimeron helicity $\eta$ as a function of time. The bimeron with $Q_{\text{v}}=+1$ is driven by $\tau_{1}$ with $\varphi=0^{\circ}$. The driving current density $j=10$ MA cm$^{-2}$.
(d) Bimeron helicity $\eta$ as a function of time. The bimeron with $Q_{\text{v}}=+1$ is driven by $\tau_{1}$ with $\varphi=180^{\circ}$. The driving current density $j=10$ MA cm$^{-2}$.
}
\label{FIG9}
\end{figure}

\begin{figure*}[t]
\centerline{\includegraphics[width=0.800\textwidth]{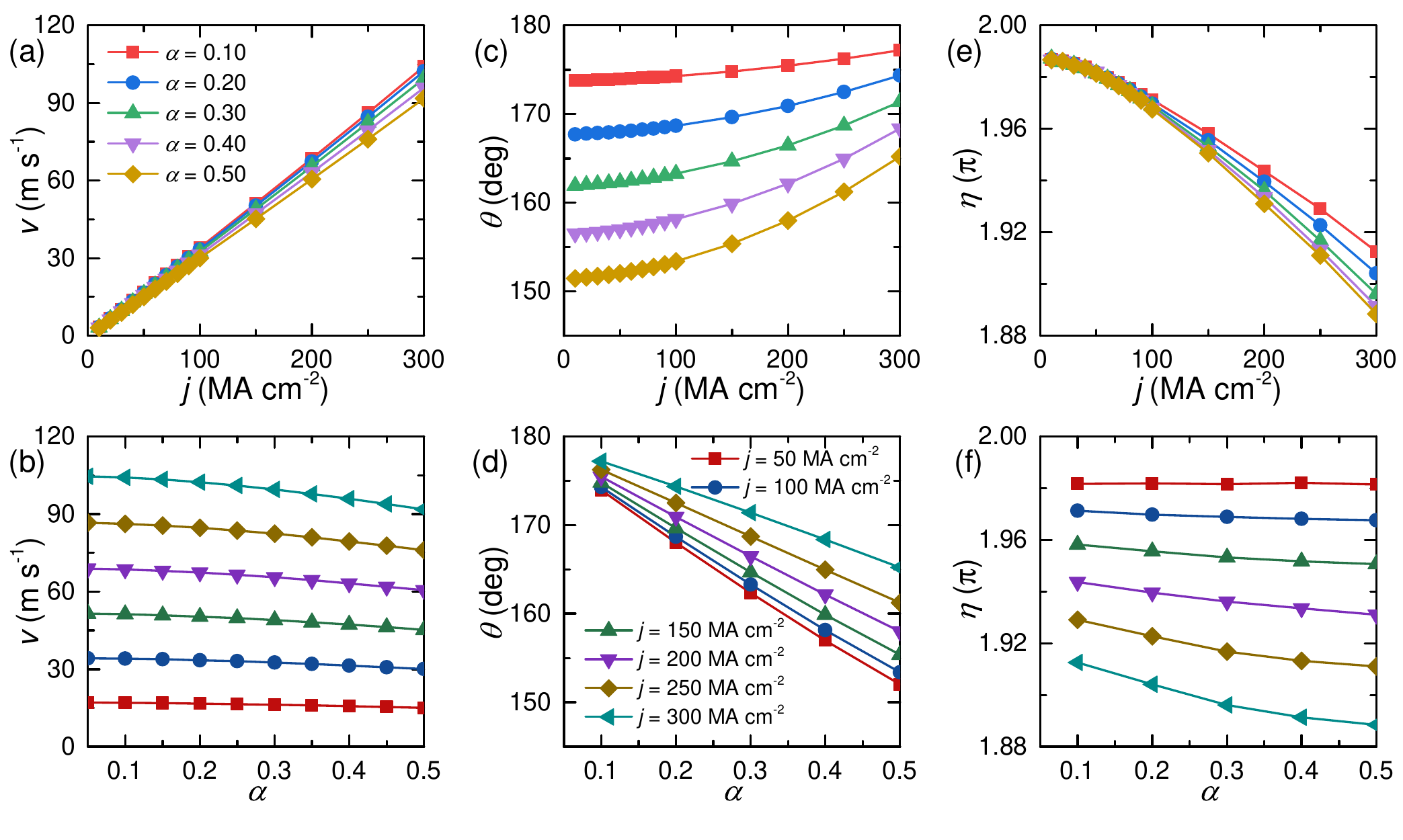}}
\caption{%
(a) Bimeron velocity $v$ as a function of driving current density for different damping parameters $\alpha$. The bimeron with $(+1,0)$ is driven by $\tau_{1}$ with $\varphi=90^{\circ}$. The bimeron shows stable linear motion.
(b) Bimeron velocity $v$ as a function of damping parameter for different driving current densities.
(c) Bimeron Hall angle $\theta$ as a function of driving current density for different damping parameters.
(d) Bimeron Hall angle $\theta$ as a function of damping parameter for different driving current densities.
(e) Bimeron helicity $\eta$ as a function of driving current density for different damping parameters.
(f) Bimeron helicity $\eta$ as a function of damping parameter for different driving current densities.
The labels for curves with light and dark colors are given in (a) and (d), respectively.
}
\label{FIG10}
\end{figure*}

\begin{figure*}[t]
\centerline{\includegraphics[width=0.800\textwidth]{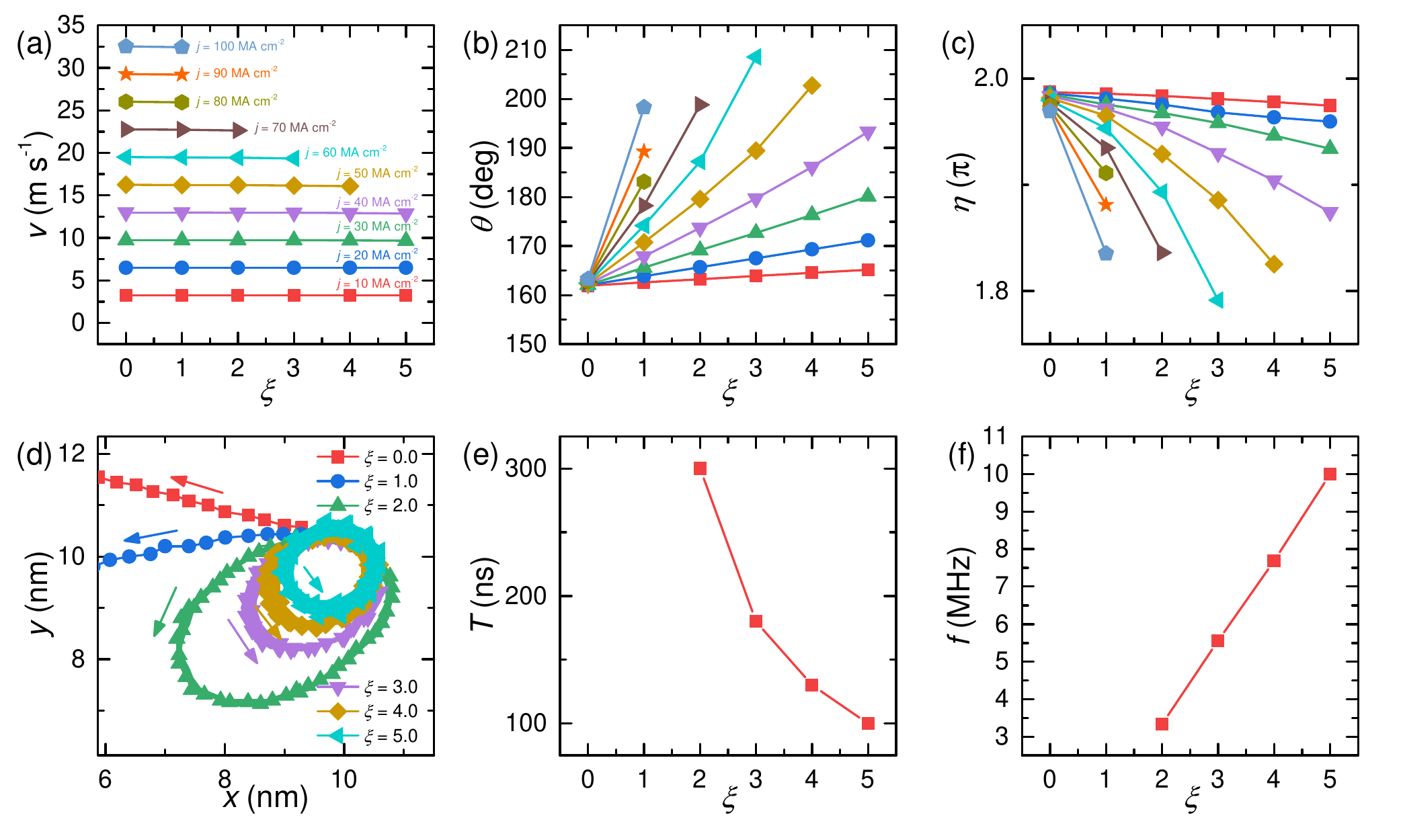}}
\caption{%
(a) Bimeron velocity $v$ as a function of fieldlike torque strength $\xi$ for different driving current densities. The bimeron with $(+1,0)$ is driven by $\tau_{1}$ and $\tau_{2}$ with $\varphi=90^{\circ}$. Only the velocity of linear motion is given.
(b) Bimeron Hall angle $\theta$ as a function of fieldlike torque strength $\xi$ for different driving current densities, corresponding to (a).
(c) Bimeron helicity $\eta$ as a function of fieldlike torque strength $\xi$ for different driving current densities, corresponding to (a).
(d) Trajectories of bimeron driven by $\tau_{1}$ and $\tau_{2}$ with $\varphi=90^{\circ}$ for different fieldlike torque strength $\xi$ at $j=100$ MA cm$^{-2}$. The dynamics of the bimeron with $(+1,0)$ changes from linear motion to elliptical motion with rotation when $\xi\geq 1$. Once the elliptical motion of bimeron is induced, its velocity $v$ and helicity $\eta$ vary with time.
(e) Period of bimeron rotation $T$ as a function of fieldlike torque strength $\xi$, corresponding to (d).
(f) Frequency of bimeron rotation $f$ as a function of fieldlike torque strength $\xi$, corresponding to (d).
}
\label{FIG11}
\end{figure*}

\begin{figure}[t]
\centerline{\includegraphics[width=0.485\textwidth]{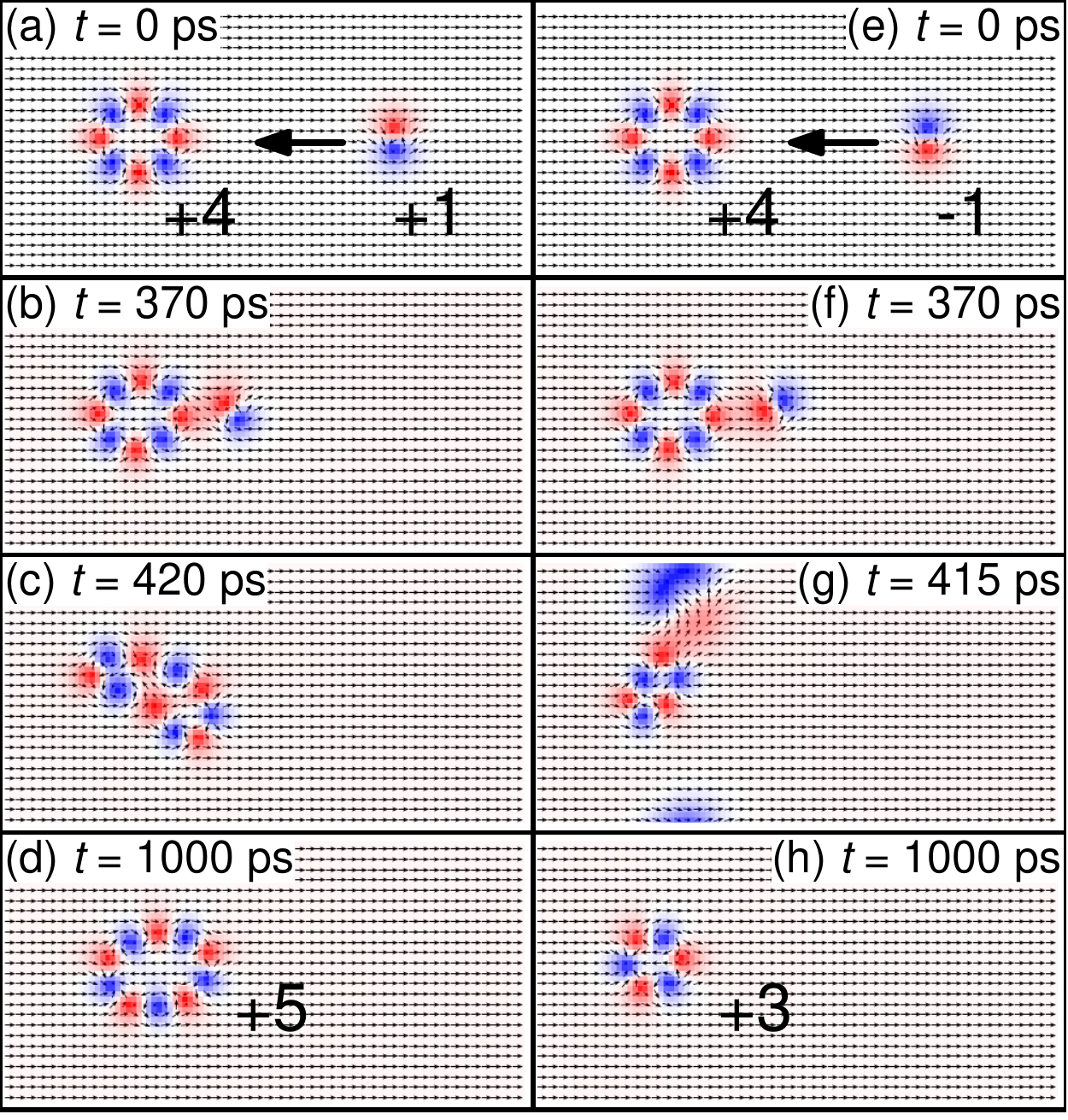}}
\caption{%
[(a)-(d)] Current-induced collision and merging of an isolated bimeron with $Q_{\text{v}}=+1$ and a bimeron state with $Q_{\text{v}}=+4$ leads to the formation of a bimeron state with $Q_{\text{v}}=+5$.
[(e)-(h)] Current-induced collision and merging of an isolated antibimeron with $Q_{\text{v}}=-1$ and a bimeron state with $Q_{\text{v}}=+4$ leads to the formation of a bimeron state with $Q_{\text{v}}=+3$.
The out-of-plane component of magnetization ($m_{z}$) is color coded: blue is into the plane, red is out of the plane, white is in-plane.
Here, $j=100$ MA cm$^{-2}$, $\varphi=90^{\circ}$, $\xi=0$, $\alpha=0.1$. Note that the dampinglike torque $\tau_{1}$ cannot drive the clusterlike bimeron state with $\left|Q_{\text{v}}\right|>1$ into motion.
}
\label{FIG12}
\end{figure}

As mentioned above, an isolated bimeron with $Q_{\text{v}}=\pm 1$ has a pair of opposite out-of-plane cores. In Fig.~\ref{FIG2}(b), it can be seen that the number of out-of-plane core pair is equal to $\left|Q_{\text{v}}\right|$. The size of the out-of-plane core is independent of $Q_{\text{v}}$, while the size of the bimeron state increases with $Q_{\text{v}}$.
We note that the bimeron state with $\left|Q_{\text{v}}\right|>1$ can be seen as a circular chain of bimerons with $Q_{\text{v}}=\pm 1$, indicating the cluster feature of high-order bimerons with $\left|Q_{\text{v}}\right|>1$. Namely, the isolated bimerons with $Q_{\text{v}}=\pm 1$ serve as the fundamental quasi-particles, which can form high-order clusterlike bimeron states.
As the total energy of a bimeron state with $\left|Q_{\text{v}}\right|=n$ is smaller than the energy sum of $n$ isolated bimeron with $Q_{\text{v}}=\pm 1$, it is envisioned that isolated bimerons with $Q_{\text{v}}=\pm 1$ are inclined to form high-order clusterlike bimeron states in the frustrated in-plane magnetic system. This feature can be seen from the metastability diagram simulated based on randomly distributed spin configurations (see Supplementary Figs.~\blue{7}-\blue{12})~\cite{SM}, where most relaxed bimeron states exist in the form of high-order clusterlike states.
It is noteworthy that this property of bimerons is in a stark contrast to magnetic skyrmions~\cite{Xichao_NCOMMS2017}. The high-order skyrmion states with $\left|Q_{\text{v}}\right|>1$ are usually unstable in either frustrated or common FM systems.

We continue to study the relaxed isolated bimerons with $Q_{\text{v}}=\pm 1$ and different values of helicity $\eta$, as shown in Fig.~\ref{FIG4}.
It can be seen that both the in-plane ($m_y$) and out-of-plane ($m_z$) spin configurations of the bimeron structure are controlled by $\eta$, which can be explained by Eq.~(\ref{eq:Q_eta}).
Figure~\ref{FIG5} shows the total energy of the relaxed isolated bimeron with $Q_{\text{v}}=\pm 1$ for $\eta=0, \pi/2, \pi, 3\pi/2$. Similar to the frustrated skyrmion with $Q_{\text{v}}=\pm 1$ (see Ref.~\onlinecite{Xichao_NCOMMS2017}), the total energy of an isolated bimeron with $Q_{\text{v}}=\pm 1$ depends on its helicity number $\eta$. Namely, the total energies of bimerons with $(\pm 1, \pi/2)$ and $(\pm 1, 3\pi/2)$ are larger than that with $(\pm 1, 0)$ and $(\pm 1, \pi)$.
It should be pointed out that all energy terms of the frustrated bimeron with $Q_{\text{v}}=\pm 1$ depend on $\eta$.
However, by analyzing the energy difference between the isolated bimeron with $\eta=0, \pi$ and the isolated bimeron with $\eta=\pi/2, 3\pi/2$ (see Fig.~\ref{FIG6}), it is found that the energy differences of the NN exchange, NNN exchange, NNNN exchange, and magnetic anisotropy almost cancel each other. The total energy difference is mainly contributed by the DDI energy difference, leading to energetically favorable bimeron structures with $(1,0)$ and $(1,\pi)$. 

\subsection{Current-driven dynamics of frustrated bimerons}
\label{se:DPB}

The current-driven dynamics of bimerons is important for the design and development of future spintronic applications, in which bimerons are movable information carriers.
Here we first study the dynamics of a relaxed isolated bimeron with $(+1,0)$ driven by the dampinglike torque $\tau_{1}$ (i.e., $\xi=0$).
Note that we limit our analysis and discussion to the isolated bimeron with $Q_{\text{v}}=+1$ in this section.
The dynamics of bimeron states with higher $Q_{\text{v}}$ will be discussed later.

As the initial state, an isolated bimeron with $(+1,0)$ is relaxed at the center of the FM monolayer with PBC. Note that $\eta=0$ is identical to $\eta=2\pi$, i.e., $(+1,0)=(+1,2\pi)$. Such a relaxed bimeron has a lower energy than that with $\eta=\pi/2$ or $\eta=3\pi/2$ (see Sec.~\ref{se:SPB}).
We assume that $J_{2}=-0.8$, $J_{3}=-0.6$, $K=0.02$, and $\alpha=0.3$ as default parameters. Other parameters are given in the method section (see Sec.~\ref{se:Methods}).
We define the angle between the spin polarization direction and the $+x$ direction in the $x-y$ plane as $\varphi$, and we first focus on the situation that the spin current is polarized along the $+y$ (i.e., $\varphi=90^{\circ}$) or $-y$ (i.e., $\varphi=270^{\circ}$) direction. Namely, the spin polarization is perpendicular to the easy axis of anisotropy. This spin polarization configuration can be realized by harnessing the spin Hall effect~\cite{Finocchio_JPD2016,Kang_PIEEE2016,Wanjun_PHYSREP2017,Fert_NATREVMAT2017,ES_JAP2018,Zhang_JPCM2020,Tomasello_SREP2014}, where a charge current flows along the $\pm x$ direction in a heavy metal substrate.

When the spin current with either $\varphi=90^{\circ}$ or $\varphi=270^{\circ}$ is applied, the relaxed isolated bimeron with $Q_{\text{v}}= +1$ will be driven into linear motion with a constant velocity. The bimeron velocity $v$, bimeron Hall angle $\theta=\arctan(v_{y}/v_{x})$, and bimeron helicity $\eta$ as functions of driving current density $j$ are given in Fig.~\ref{FIG7}.
The bimeron driven by the spin current with $\varphi=90^{\circ}$ moves toward the left direction and shows a bimeron Hall angle of $\theta\sim 162^{\circ}$, while the bimeron driven by the spin current with $\varphi=270^{\circ}$ moves toward the right direction and shows a bimeron Hall angle of $\theta\sim 342^{\circ}$ [see Fig.~\ref{FIG8}(a)]. The bimeron velocity is linearly proportional to the driving current density for stable linear motion.
We note that such a steady linear motion of bimeron driven by the dampinglike torque $\tau_{1}$ can also be described by Thiele equations (see Supplementary Note 1)~\cite{SM,Thiele_PRL1973}, provided that the bimeron is a rigid object during the motion (i.e., the shape, size and helicity are fixed).

However, it should be noted that the bimeron Hall angle $\theta$ slightly increases with increasing current density, and the bimeron helicity $\eta$ slightly decreases with increasing current density.
The reason is that a large current leads to the distortion of bimeron.
In particular, when the current density is larger than $300$ MA cm$^{-2}$, the dynamic behavior of the bimeron driven by the dampinglike torque with $\varphi=270^{\circ}$ suddenly changes from linear motion to elliptical motion with rotating helicity. Both the velocity and helicity vary with time.
Figure~\ref{FIG8}(a) shows the bimeron trajectories driven by a small current density $j=40$ MA cm$^{-2}$. The corresponding velocity [Fig.~\ref{FIG8}(b)] and helicity [Fig.~\ref{FIG8}(c)] are independent of time.
Figure~\ref{FIG8}(d) shows the bimeron trajectories driven by a large current density $j=400$ MA cm$^{-2}$. It can be seen that the bimeron driven by the dampinglike torque with $\varphi=270^{\circ}$ shows a counterclockwise elliptical motion. Its velocity oscillates between $\sim 61$ m s$^{-1}$ and $\sim 127$ m s$^{-1}$ [Fig.~\ref{FIG8}(e)]. Its helicity varies between $2\pi$ and $\sim 0$, indicating a counterclockwise rotation of the bimeron. Note that when the spin polarization is aligned along the $+y$ direction (i.e., $\varphi=90^{\circ}$), the bimeron still shows a linear motion at $j=400$ MA cm$^{-2}$, of which the velocity and helicity are equal to $\sim 135$ m s$^{-1}$ and $\sim 1.9\pi$, respectively.

As the bimeron structure with $(+1,0)$ [i.e., $(+1,2\pi)$] is asymmetric with respect to the $x$ axis (i.e., the easy axis), we find that the threshold current density for the transition of bimeron dynamics is different for $\varphi=90^{\circ}$ and $\varphi=270^{\circ}$. Indeed, the bimeron driven by the dampinglike torque with $\varphi=90^{\circ}$ also changes from linear motion to counterclockwise elliptical motion when $j\geq 600$ MA cm$^{-2}$ (see Supplementary Fig.~\blue{19})~\cite{SM}.
However, one should note that the dynamics driven by large current density may be unrealistic from the point of view of experiments.
Also, the dynamics driven by large current density cannot be analyzed by simple Thiele equations due to the current-induced distortion and instability of bimeron, but rather require much more complicated analysis based on generalized Thiele equations involving many collective coordinates~\cite{Tretiakov_2008A, Tretiakov_2008B}.
Therefore, in this work, we focus on the bimeron dynamics driven by a moderate current density, i.e., $j\leq 300$ MA cm$^{-2}$.

We continue to investigate the bimeron dynamics driven by the dampinglike torque, of which the spin polarization direction is aligned along the $\pm x$ direction (i.e., $\varphi=0^{\circ}$ or $\varphi=180^{\circ}$). Namely, the spin polarization is parallel to the easy axis of anisotropy. Such a spin polarization configuration can be realized by injecting a charge current flowing along the $\pm y$ direction in a heavy metal substrate. An isolated bimeron with $(+1,0)$ is relaxed at the center of the FM monolayer with PBC as the initial state.

When the spin current with either $\varphi=0^{\circ}$ or $\varphi=180^{\circ}$ is applied, the relaxed isolated bimeron with $Q_{\text{v}}= +1$ will be driven into rotation with a specific rotation period $T$ and frequency $f$, which is in stark contrast to the case driven by the dampinglike torque with $\varphi=90^{\circ}$ or $\varphi=270^{\circ}$.
Namely, no linear motion is induced by the dampinglike torque $\tau_{1}$ with $\varphi=0^{\circ}$ or $\varphi=180^{\circ}$, which is in line with the results reported in Ref.~\onlinecite{Gobel_PRB2019} and can also be found by Thiele equations (see Supplementary Note 1)~\cite{SM}.

It is found that the bimeron rotation period $T$ decreases with the driving current density [see Fig.~\ref{FIG9}(a)]. The bimeron rotation frequency $f$ is almost linearly proportional to the driving current density [see Fig.~\ref{FIG9}(b)].
The bimeron driven by the dampinglike torque with $\varphi=0^{\circ}$ shows a counterclockwise rotation, while the bimeron driven by the dampinglike torque with $\varphi=180^{\circ}$ shows a clockwise rotation. The helicity of the rotating bimeron as a function of time is given in Fig.~\ref{FIG9}.
The time-dependent helicity numbers of rotating bimerons driven by the dampinglike torque with $\varphi=0^{\circ}$ and $\varphi=180^{\circ}$ are given in Figs.~\ref{FIG9}(c) and \ref{FIG9}(d), respectively.

From the viewpoint of applications, the isolated bimerons showing the linear motion can be used as information carriers in data storage devices, such as the racetrack-type memory. On the other hand, the isolated bimerons showing rotation or elliptical motion can be used to generate spin waves in low-damping materials, which may be used for building nano-oscillators.
In the following, we focus on the stable linear motion of an isolated bimeron and study the effects of damping parameter $\alpha$ and fieldlike torque $\tau_{2}$ on the dynamics.

The initial state is also a relaxed isolated bimeron with $(+1,0)$ at the center of the FM monolayer with PBC. The spin current with a polarization direction of $\varphi=90^{\circ}$ is applied to drive the bimeron into stable linear motion. The driving current density ranges from $10$ MA cm$^{-2}$ to $300$ MA cm$^{-2}$. Note that only the dampinglike torque $\tau_{1}$ is applied here (i.e., $\xi=0$).
Figure~\ref{FIG10}(a) shows the velocity of bimeron as a function of driving current density for different damping parameters. It can be seen that the velocity is proportional to the driving current density, while the velocity decreases with increasing damping parameter for a given driving current density [see Fig.~\ref{FIG10}(b)].
Figure~\ref{FIG10}(c) shows the bimeron Hall angle as a function of driving current density for different damping parameters. The bimeron Hall angle increases with driving current density. The dependence of bimeron Hall angle on the driving current density is significant for larger damping parameter. For a given driving current density, the bimeron Hall angle is inversely proportional to the damping parameter [see Fig.~\ref{FIG10}(d)]. Smaller driving current density leads to larger slope of the $\theta$-$\alpha$ relation.
Figure~\ref{FIG10}(e) shows the bimeron helicity as a function of driving current density for different damping parameters. The bimeron helicity slightly decreases with increasing driving current density. For a given driving current density, the bimeron helicity decreases with increasing damping parameter, which is remarkable for larger driving current density [see Fig.~\ref{FIG10}(f)].

In order to study the effect of fieldlike torque $\tau_{2}$ on the bimeron dynamics, we apply a spin current with a polarization direction of $\varphi=90^{\circ}$ to drive the bimeron with $(+1,0)$. The driving current density ranges from $10$ MA cm$^{-2}$ to $100$ MA cm$^{-2}$. Both the dampinglike $\tau_{1}$ and fieldlike $\tau_{2}$ torques are applied.
It is found that the bimeron could be driven into linear motion or elliptical motion with rotation in the presence of fieldlike torque $\tau_{2}$, which depends on the strength of the fieldlike torque $\xi$ as well as the driving current density $j$. We also find that the threshold value of the fieldlike torque strength $\xi$, beyond which the elliptical motion is induced, decreases with increasing driving current density.

For the stable linear motion of bimeron driven by both dampinglike and fieldlike torques, the velocity as a function of the fieldlike torque strength $\xi$ for different driving current densities is given in Fig.~\ref{FIG11}(a). Note that only the velocity of stable linear motion is given, while the velocity of elliptical motion varies with time [see Fig.~\ref{FIG8}(e)] and thus, is not given in Fig.~\ref{FIG11}(a). It can be seen that the velocity of stable linear motion is independent of the fieldlike torque strength $\xi$ for a given driving current density.
However, for the linear motion of bimeron driven by a given current density, the bimeron Hall angle $\theta$ increases with increasing fieldlike torque strength $\xi$, and the slope of the $\theta$-$\xi$ relation increases with the driving current density, as shown in Fig.~\ref{FIG11}(b).
Figure~\ref{FIG11}(c) shows the bimeron helicity as a function of the fieldlike torque strength $\xi$ for different driving current densities. The bimeron helicity decreases with increasing fieldlike torque strength $\xi$ for a given driving current density. Also, the bimeron helicity decreases faster with increasing fieldlike torque strength $\xi$ when the driving current density is larger.
The dependence of bimeron Hall angle and helicity on the fieldlike torque strength $\xi$ indicates that the fieldlike torque $\tau_{2}$ can deform the bimeron.

For a given driving current density, the bimeron will be driven into elliptical motion accompanied by rotation when the fieldlike torque strength $\xi$ is larger than a certain threshold. Figure~\ref{FIG11}(d) shows the trajectories of bimeron driven by the spin current of $j=100$ MA cm$^{-2}$ for different strength $\xi$ of the fieldlike torque. The current-driven bimeron shows linear motion when $\xi\leq 1$, while it shows elliptical motion when $\xi\geq 2$. Similar to the elliptical motion induced by a large dampinglike torque $\tau_{1}$ with $\varphi=0^{\circ}$ or $\varphi=180^{\circ}$, the period of the bimeron rotation decreases with the fieldlike torque strength $\xi$ [see Fig.~\ref{FIG11}(e)], and the bimeron rotation frequency is linearly proportional to the fieldlike torque strength $\xi$ [see Fig.~\ref{FIG11}(f)].

Although the isolated bimeron with $Q_{\text{v}}=+1$ can be driven into linear motion by the dampinglike torque $\tau_{1}$ with a moderate driving current density and a spin polarization aligned along the $\pm y$ direction (i.e., $\varphi=90^{\circ}$ or $\varphi=270^{\circ}$), we note that motion direction of the isolated bimeron also depends on its helicity $\eta$ (see Supplementary Fig.~\blue{20} and Supplementary Note 1)~\cite{SM}, which can also be discerned from the elliptical motion dynamics [cf. Figs.~\ref{FIG8}(d) and \ref{FIG8}(f)].
However, since the clusterlike bimeron state with a higher vorticity number $\left|Q_{\text{v}}\right|=n$ ($n\geq 2$) can be regarded as a composite of $n$ isolated bimerons with $Q_{\text{v}}=\pm 1$ with varied helicity numbers [see Fig.~\ref{FIG2}(b)], the helicity-dependent driving forces acting on each bimeron with $Q_{\text{v}}=\pm 1$ could cancel each other.
As a result, the clusterlike bimeron state with a higher vorticity number cannot be driven into directional motion by the dampinglike torque $\tau_{1}$ with $\varphi=90^{\circ}$ or $\varphi=270^{\circ}$.
Such a phenomenon can also be found by Thiele equations (see Supplementary Note 1)~\cite{SM}, where the driving-force-related tensor is equal to zero for spin configurations of bimeron states with $\left|Q_{\text{v}}\right|>1$, indicating the total driving force provided by the dampinglike torque $\tau_{1}$ on bimeron states with $\left|Q_{\text{v}}\right|>1$ equals zero.

Therefore, we find it is possible to bombard a clusterlike bimeron state having higher $Q_{\text{v}}$ with an isolated bimeron having $Q_{\text{v}}=\pm 1$.
Figure~\ref{FIG12} shows the snapshots of the computational experiment on the collision and merging of bimeron states.
We put a relaxed clusterlike bimeron state with $Q_{\text{v}}=+4$ at the left side of the sample, and a relaxed isolated bimeron with $Q_{\text{v}}=+1$ at the right side of the sample. Then, we apply a spin current with $j=100$ MA cm$^{-2}$ and $\varphi=90^{\circ}$ to the sample. Note that we only use the dampinglike torque $\tau_{1}$ for the sake of simplicity. Once the driving current is applied, it is found that the isolated bimeron with $Q_{\text{v}}=+1$ moves toward the left direction, while the clusterlike bimeron state with $Q_{\text{v}}=+4$ keeps still. Hence, the collision and merging between the two bimeron states lead to the creation of a clusterlike bimeron state with $Q_{\text{v}}=+5$. In a similar way, if we put an isolated antibimeron with $Q_{\text{v}}=-1$ at the right side of the sample, the collision and merging between the isolated antibimeron with $Q_{\text{v}}=-1$ and the bimeron state with $Q_{\text{v}}=+4$ lead to the creation of a clusterlike bimeron state with $Q_{\text{v}}=+3$.
It can be seen that the topological charge of the whole system is conserved during the collision and merging of bimeron states.

As discussed in Sec.~\ref{se:SPB} [see Fig.~\ref{FIG2}(a)], due to the energy minimization of the bimeron system, the isolated bimeron with $Q_{\text{v}}= +1$ can be merged with a bimeron state with $Q_{\text{v}}= n$, forming a bimeron state with a higher $Q_{\text{v}}= n+1$. Also, the merging of an isolated antibimeron with $Q_{\text{v}}= -1$ and a bimeron state with $Q_{\text{v}}= n$ will lead to the formation of a bimeron state with lower $Q_{\text{v}}= n-1$. The total energy of the bimeron system is reduced after the collision and merging.
Such a process is verified by the computational experiment on the collision and merging of bimeron states. In principle, it is possible to change the topological charge of a bimeron state by the bimeron collision and merging, which is worth investigating in future experiments.

\section{Conclusion}
\label{se:Conclusion}

In conclusion, we have numerically studied the static and dynamic properties of bimerons in a ferromagnetic monolayer system with competing exchange interactions.
A candidate material for such a system could be the low-dimensional compound Pb$_{2}$VO(PO$_{4}$)$_2$~\cite{Kaul_JMMM2004}, which has a frustrated square lattice with small ferromagnetic nearest-neighbor exchange interaction and strong antiferromagnetic next-nearest-neighbor exchange interaction.

We considered the long-range dipole-dipole interaction in our simulations, which has not been considered in other studies~\cite{Kharkov_PRL2017,Gobel_PRB2019}.
Although the long-range dipole-dipole interaction is small in in-plane monolayer system, it can lead to the helicity dependence of the isolated bimeron energy for $Q_{\text{v}}=\pm 1$, which is essential for the helicity-dependent bimeron dynamics.
The bimeron states can be stabilized by the exchange frustration both in the presence and absence of in-plane magnetic anisotropy, and the presence of in-plane magnetic anisotropy could result in a smaller bimeron.

In Sec.~\ref{se:SPB}, we studied the energies of bimeron states in the presence of finite in-plane magnetic anisotropy.
The total energy of a bimeron state increases with its vorticity number $Q_{\text{v}}$. On the other hand, the total energy of a bimeron with $Q_{\text{v}}=\pm 1$ depends on its helicity $\eta$, for example, the states with $\eta=0$ and $\eta=\pi$ have a lower energy compared to that with $\eta=\pi/2$ and $\eta=3\pi/2$.
Although the contribution of the dipole-dipole interaction energy to the total energy of a bimeron is small, the energy dependence on the helicity is induced by the dipole-dipole interaction, similar to the case of frustrated skyrmions~\cite{Xichao_NCOMMS2017}.
We note that the bimeron state with a higher vorticity number $\left|Q_{\text{v}}\right|>1$ can be seen as a clusterlike state formed by a circular chain of bimerons with $Q_{\text{v}}=\pm 1$. In this work, however, we mainly focused on the isolated bimeron with $Q_{\text{v}}=\pm 1$.

In Sec.~\ref{se:DPB}, we studied the dynamics of a single isolated bimeron with $Q_{\text{v}}=+1$ driven by the spin-orbit torques, including the dampinglike and fieldlike torques.
For the bimeron driven by the dampinglike torque, the current-induced dynamics depends on the spin polarization direction $\boldsymbol{p}$ with respect to the easy-axis direction.
Note that the easy axis of in-plane magnetic anisotropy is aligned along the $x$ axis in this work.

When the spin polarization is perpendicular to the easy axis (i.e., $\varphi=90^{\circ}$ or $\varphi=270^{\circ}$), the bimeron with $Q_{\text{v}}=+1$ can be driven into stable linear motion for a moderate driving current density. The velocity and helicity for the stable linear motion are independent of time. For the linear motion of a bimeron, the bimeron Hall angle is inversely proportional to the damping parameter.
However, when the driving current density is larger than a certain threshold, the bimeron can be driven into elliptical motion accompanied by rotation.
Both the velocity and helicity vary with time during the elliptical motion.
We note that the dynamics of a single isolated antibimeron with $Q_{\text{v}}=-1$ is similar to the case of $Q_{\text{v}}=+1$ (see Supplementary Fig.~\blue{21})~\cite{SM}.

When the spin polarization is parallel to the easy axis (i.e., $\varphi=0^{\circ}$ or $\varphi=180^{\circ}$), the bimeron with $Q_{\text{v}}=+1$ will be driven into rotation only. The rotating bimeron has a constant frequency that is almost linearly proportional to the driving current density.

For the bimeron with $Q_{\text{v}}=+1$ driven by both the dampinglike and fieldlike torques with a spin polarization perpendicular to the easy axis (i.e., $\varphi=90^{\circ}$), the bimeron dynamics will change from the linear motion to the elliptical motion when the fieldlike torque strength is larger than a certain threshold.
In the case of the linear motion (i.e., $\xi$ is small), the bimeron Hall angle and bimeron helicity can be changed by the fieldlike torque, especially for the bimeron driven by a large current density.
In the case of the elliptical motion induced by the fieldlike torque (i.e., $\xi$ is large), the frequency of the bimeron rotation is linearly proportional to the strength of the fieldlike torque.

In Sec.~\ref{se:DPB}, we also performed a computational experiment on the current-driven collision and merging of bimeron states, namely, we used an isolated bimeron with $Q_{\text{v}}=\pm 1$ to bombard a clusterlike bimeron state with a higher topological charge. The collision and merging between the isolated bimeron and clusterlike bimeron state lead to the formation of a new bimeron state. The topological charge of the whole system during the bimeron collision and merging process is conserved, while the total energy of the whole system is reduced. The bimeron collision and merging experiment shows the possibility that one can create a bimeron state with a higher or lower topological charge by triggering the collision and merging of isolated bimerons with $Q_{\text{v}}=\pm 1$.

Last, we point out both the advantages and disadvantages of frustrated bimerons compared to typical skyrmions.
First, from the point of view of spintronic applications, both frustrated bimerons and skyrmions are useful for information storage.
Namely, the frustrated bimerons and skyrmions~\cite{Gobel_PRB2019,Lin_PRB2016A,Leonov_NCOMMS2017,Xichao_NCOMMS2017,Xia_PRApplied2019} showing the directional linear motion can be used as building blocks for the racetrack-type memory, where bimerons and skyrmions serve as robust and movable information carriers.
In particular, as mentioned in Ref.~\onlinecite{Gobel_PRB2019}, the smaller stray fields in in-plane magnetized nanotracks allow for a denser array in three dimensions, and thus allow for a higher packing density and storage density of bimeron-based racetrack-type memory.
On the other hand, the elliptical motion of frustrated bimerons and the circular motion of frustrated skyrmions~\cite{Lin_PRB2016A,Leonov_NCOMMS2015,Xichao_NCOMMS2017,Xia_PRApplied2019} can be used to build spintronic devices like nano-oscillators.
Also, the rotating frustrated bimeron can be used as a spin-wave source in nanoscale magnonic devices.

Indeed, compared to typical skyrmions, frustrated bimerons also have some unique properties, which may lead to limitations in spintronic applications.
For example, the isolated frustrated bimerons with $Q_{\text{v}}=\pm 1$ are inclined to form high-order clusterlike bimeron states, however, the high-order bimeron states with $\left|Q_{\text{v}}\right|>1$ cannot be driven into linear motion by the dampinglike torque. Consequently, it is unrealistic to use the high-order bimeron states with $\left|Q_{\text{v}}\right|>1$ in bimeron shift memory devices.
Nevertheless, it is still envisioned that the high-order clusterlike bimeron states can be used as information carriers in multi-state magnetic memory devices.
We believe that our results are useful for understanding the bimeron physics in frustrated magnets, and could provide guidelines for building bimeron-based spintronic devices.

\begin{acknowledgments}
X.Z. was supported by the Guangdong Basic and Applied Basic Research Foundation (Grant No. 2019A1515110713), and the Presidential Postdoctoral Fellowship of The Chinese University of Hong Kong, Shenzhen (CUHKSZ).
M.E. acknowledges the support by the Grants-in-Aid for Scientific Research from JSPS KAKENHI (Grant Nos. JP18H03676, JP17K05490, and JP15H05854) and the support by CREST, JST (Grant Nos. JPMJCR16F1 and JPMJCR1874).
O.A.T. acknowledges the support by the Australian Research Council (Grant No. DP200101027), the Cooperative Research Project Program at the Research Institute of Electrical Communication, Tohoku University, and by UNSW Science International Seed Grant.
X.L. acknowledges the support by the Grants-in-Aid for Scientific Research from JSPS KAKENHI (Grant Nos. 17K19074, 26600041, and 22360122).
G.Z. acknowledges the support by the National Natural Science Foundation of China (Grant Nos. 51771127, 51571126, and 51772004), the Scientific Research Fund of Sichuan Provincial Education Department (Grant Nos. 18TD0010 and 16CZ0006).
Y.Z. acknowledges the support by the President's Fund of CUHKSZ, Longgang Key Laboratory of Applied Spintronics, National Natural Science Foundation of China (Grant Nos. 11974298 and 61961136006), Shenzhen Fundamental Research Fund (Grant No. JCYJ20170410171958839), and Shenzhen Peacock Group Plan (Grant No. KQTD20180413181702403).
\end{acknowledgments}





\begin{thebibliography}{99}

\bibitem{Nagaosa_NNANO2013} N. Nagaosa and Y. Tokura, Topological properties and dynamics of magnetic skyrmions, Nat. Nanotech. \textbf{8}, 899 (2013).

\bibitem{Wiesendanger_NATREVMAT2016} R. Wiesendanger, Nanoscale magnetic skyrmions in metallic films and multilayers: a new twist for spintronics, Nat. Rev. Mats. \textbf{1}, 16044 (2016).

\bibitem{Finocchio_JPD2016} G. Finocchio, F. B{\"u}ttner, R. Tomasello, M. Carpentieri, and M. Kl{\"a}ui, Magnetic skyrmions: from fundamental to applications, J. Phys. D: Appl. Phys. \textbf{49}, 423001 (2016).

\bibitem{Kang_PIEEE2016} W. Kang, Y. Huang, X. Zhang, Y. Zhou, and W. Zhao, Skyrmion-electronics: an overview and outlook, Proc. IEEE \textbf{104}, 2040 (2016).

\bibitem{Kanazawa_AM2017} N. Kanazawa, S. Seki, and Y. Tokura, Noncentrosymmetric magnets hosting magnetic skyrmions, Adv. Mater. \textbf{29}, 1603227 (2017).

\bibitem{Wanjun_PHYSREP2017} W. Jiang, G. Chen, K. Liu, J. Zang, S. G. Velthuiste, and A. Hoffmann, Skyrmions in magnetic multilayers, Phys. Rep. \textbf{704}, 1 (2017).

\bibitem{Fert_NATREVMAT2017} A. Fert, N. Reyren, and V. Cros, Magnetic skyrmions: advances in physics and potential applications, Nat. Rev. Mater. \textbf{2}, 17031 (2017).

\bibitem{Zhou_NSR2018} Y. Zhou, Magnetic skyrmions: intriguing physics and new spintronic device concepts, Natl. Sci. Rev. \textbf{6}, 210 (2019).

\bibitem{ES_JAP2018} K. Everschor-Sitte, J. Masell,  R. M. Reeve, and M. Kl\"{a}ui, Perspective: magnetic skyrmions-overview of recent progress in an active research field, J. Appl. Phys. \textbf{124}, 240901 (2018).

\bibitem{Zhang_JPCM2020} X. Zhang, Y. Zhou, K. M. Song, T.-E. Park, J. Xia, M. Ezawa, X. Liu, W. Zhao, G. Zhao, and S. Woo,
Skyrmion-electronics: writing, deleting, reading and processing magnetic skyrmions toward spintronic applications, J. Phys. Condens. Matter \textbf{32}, 143001 (2020).


\bibitem{Bogdanov_1989} A. N. Bogdanov and D. A. Yablonskii, Thermodynamically stable `vortices' in magnetically ordered crystals. The mixed state of magnets. Sov. Phys. JETP \textbf{68}, 101 (1989).

\bibitem{Roszler_NATURE2006} U. K. R{\"o}{\ss}ler, A. N. Bogdanov, and C. Pfleiderer, Spontaneous skyrmion ground states in magnetic metals, Nature \textbf{442}, 797 (2006).


\bibitem{Muhlbauer_SCIENCE2009} S. M{\"u}hlbauer, B. Binz, F. Jonietz, C. Pfleiderer, A. Rosch, A. Neubauer, R. Georgii, and P. B{\"o}ni, Skyrmion lattice in a chiral magnet, Science \textbf{323}, 915 (2009).


\bibitem{Lin_PRB2013} S.-Z. Lin, C. Reichhardt, C. D. Batista, and A. Saxena, Particle model for skyrmions in metallic chiral magnets: dynamics, pinning, and creep, Phys. Rev. B \textbf{87}, 214419 (2013).

\bibitem{Reichhardt_2017} C. Reichhardt and C. J. Olson Reichhardt, Depinning and nonequilibrium dynamic phases of particle assemblies driven over random and ordered substrates: a review, Rep. Prog. Phys. \textbf{80}, 026501 (2017).


\bibitem{Yu_NATURE2010} X. Z. Yu, Y. Onose, N. Kanazawa, J. H. Park, J. H. Han, Y. Matsui, N. Nagaosa, and Y. Tokura, Real-space observation of a two-dimensional skyrmion crystal, Nature \textbf{465}, 901 (2010).

\bibitem{Du_NCOMMS2015} H. Du, R. Che, L. Kong, X. Zhao, C. Jin, C. Wang, J. Yang, W. Ning, R. Li, C. Jin, X. Chen, J. Zang, Y. Zhan, and M. Tian, Edge-mediated skyrmion chain and its collective dynamics in a confined geometry, Nat. Commun. \textbf{6}, 8504 (2015).

\bibitem{Jiang_SCIENCE2015} W. Jiang, P. Upadhyaya, W. Zhang, G. Yu, M. Benjamin Jungfleisch, F. Y. Fradin, J. E. Pearson, Y. Tserkovnyak, K. L. Wang, O. Heinonen, S. G. E. te Velthuis, A. Hoffmann, Blowing magnetic skyrmion bubbles, Science \textbf{349}, 283 (2015).

\bibitem{Woo_NMATER2016} S. Woo, K. Litzius, B. Kruger, M.-Y. Im, L. Caretta, K. Richter, M. Mann, A. Krone, R. M. Reeve, M. Weigand, P. Agrawal, I. Lemesh, M.-A. Mawass, P. Fischer, M. Klaui, and G. S. D. Beach, Observation of room-temperature magnetic skyrmions and their current-driven dynamics in ultrathin metallic ferromagnets, Nat. Mater. \textbf{15}, 501 (2016).

\bibitem{MoreauLuchaire_NNANO2016} C. Moreau-Luchaire, C. Moutafis, N. Reyren, J. Sampaio, C. A. F. Vaz, N. Van Horne, K. Bouzehouane, K. Garcia, C. Deranlot, P. Warnicke, P. Wohlh\"{u}ter, J.-M. George, M. Weigand, J. Raabe, V. Cros, and A. Fert, Additive interfacial chiral interaction in multilayers for stabilization of small individual skyrmions at room temperature, Nat. Nanotech. \textbf{11}, 444 (2016).

\bibitem{Matsumoto_SA2016} T. Matsumoto, Y.-G. So, Y. Kohno, H. Sawada, Y. Ikuhara, and N. Shibata, Direct observation of $\Sigma$7 domain boundary core structure in magnetic skyrmion lattice, Sci. Adv. \textbf{2}, e1501280 (2016).

\bibitem{Wanjun_NPHYS2017} W. Jiang, X. Zhang, G. Yu, W. Zhang, X. Wang, M. Benjamin Jungfleisch, J. E. Pearson, X. Cheng, O. Heinonen, K. L. Wang, Y. Zhou, A. Hoffmann, and S. G. E. Velthuiste, Direct observation of the skyrmion Hall effect, Nat. Phys. \textbf{13}, 162 (2017).

\bibitem{Litzius_NPHYS2017} K. Litzius, I. Lemesh, B. Kruger, P. Bassirian, L. Caretta, K. Richter, F. Buttner, K. Sato, O. A. Tretiakov, J. Forster, R. M. Reeve, M. Weigand, I. Bykova, H. Stoll, G. Schutz, G. S. D. Beach, and M. Klaui, Skyrmion Hall effect revealed by direct time-resolved X-ray microscopy, Nat. Phys. \textbf{13}, 170 (2017).

\bibitem{Pollard_NCOMMS2017} S. D. Pollard, J. A. Garlow, J. Yu, Z. Wang, Y. Zhu, and H. Yang, Observation of stable Néel skyrmions in cobalt/palladium multilayers with Lorentz transmission electron microscopy, Nat. Commun. \textbf{8}, 14761 (2017).

\bibitem{Hrabec_NC2017} A. Hrabec, J. Sampaio, M. Belmeguenai, I. Gross, R. Weil, S. M. Ch{\'e}rif, A. Stashkevich, V. Jacques, A. Thiaville, and S. Rohart, Current-induced skyrmion generation and dynamics in symmetric bilayers, Nat. Commun. \textbf{8}, 15765 (2017).

\bibitem{Woo_NC2017} S. Woo, K. M. Song, H.-S. Han, M.-S. Jung, M.-Y. Im, K.-S. Lee, K. S. Song, P. Fischer, J.-I. Hong, J. W. Choi, B.-C. Min, H. C. Koo, and J. Chang, Spin-orbit torque-driven skyrmion dynamics revealed by time-resolved X-ray microscopy, Nat. Commun. \textbf{8}, 15573 (2017).

\bibitem{Woo_NatElect2018} S. Woo, K. M. Song, X. Zhang, M. Ezawa, Y. Zhou, X. Liu, M. Weigand, S. Finizio, J. Raabe, M.-C. Park, K.-Y. Lee, J. W. Choi, B.-C. Min, H. C. Koo, and J. Chang, Deterministic creation and deletion of a single magnetic skyrmion observed by direct time-resolved X-ray microscopy, Nat. Electron. \textbf{1}, 288 (2018).

\bibitem{Woo_NC2018} S. Woo, K. M. Song, X. Zhang, Y. Zhou, M. Ezawa, X. Liu, S. Finizio, J. Raabe, N. J. Lee, S.-I. Kim, S.-Y. Park, Y. Kim, J.-Y. Kim, D. Lee, O. Lee, J. W. Choi, B.-C. Min, H. C. Koo, and J. Chang, Current-driven dynamics and inhibition of the skyrmion Hall effect of ferrimagnetic skyrmions in GdFeCo films, Nat. Commun. \textbf{9}, 959 (2018).

\bibitem{Ma_NL2019} C. Ma, X. Zhang, J. Xia, M. Ezawa, W. Jiang, T. Ono, S. N. Piramanayagam, A. Morisako, Y. Zhou, and X. Liu, Electric field-induced creation and directional motion of domain walls and skyrmion bubbles, Nano Lett. \textbf{19}, 353 (2019).

\bibitem{Nozaki_APL2019} T. Nozaki, Y. Jibiki, M. Goto, E. Tamura, T. Nozaki, H. Kubota, A. Fukushima, S. Yuasa, and Y. Suzuki, Brownian motion of skyrmion bubbles and its control by voltage applications, Appl. Phys. Lett. \textbf{114}, 012402 (2019).


\bibitem{Sampaio_NNANO2013} J. Sampaio, V. Cros, S. Rohart, A. Thiaville, and A. Fert, Nucleation, stability and current-induced motion of isolated magnetic skyrmions in nanostructures, Nat. Nanotech. \textbf{8}, 839 (2013).

\bibitem{Tomasello_SREP2014} R. Tomasello, E. Martinez, R. Zivieri, L. Torres, M. Carpentieri, and G. Finocchio, A strategy for the design of skyrmion racetrack memories, Sci. Rep. \textbf{4}, 6784 (2014).

\bibitem{Xichao_SREP2015B} X. Zhang, M. Ezawa, and Y. Zhou, Magnetic skyrmion logic gates: conversion, duplication and merging of skyrmions, Sci. Rep. \textbf{5}, 9400 (2015).

\bibitem{Senfu_NJP2015} S. Zhang, J. Wang, Q. Zheng, Q. Zhu, X. Liu, S. Chen, C. Jin, Q. Liu, C. Jia, and D. Xue, Current-induced magnetic skyrmions oscillator, New J. Phys. \textbf{17}, 023061 (2015).

\bibitem{Xichao_NCOMMS2016} X. Zhang, Y. Zhou, and M. Ezawa, Magnetic bilayer-skyrmions without skyrmion Hall effect, Nat. Commun. \textbf{7}, 10293 (2016).

\bibitem{Zhang_PRB2016} X. Zhang, M. Ezawa, and Y. Zhou, Thermally stable magnetic skyrmions in multilayer synthetic antiferromagnetic racetracks, Phys. Rev. B \textbf{94}, 064406 (2016).

\bibitem{Bourianoff_AIP2016} G. Bourianoff, D. Pinna, M. Sitte, and K. Everschor-Sitte, Potential implementation of reservoir computing models based on magnetic skyrmions, AIP Adv. \textbf{8}, 055602 (2018).

\bibitem{Tomasello_JPD2017} R. Tomasello, V. Puliafito, E. Martinez, A. Manchon, M. Ricci, M. Carpentieri, and G. Finocchio, Performance of synthetic antiferromagnetic racetrack memory: domain wall versus skyrmion, J. Phys. D: Appl. Phys. \textbf{50}, 325302 (2017).

\bibitem{Muller_NJP2017} J. M{\"u}ller, Magnetic skyrmions on a two-lane racetrack, New J. Phys. \textbf{19}, 025002 (2017).

\bibitem{Yangqi_NANO2017} Y. Huang, W. Kang, X. Zhang, Y. Zhou, and W. Zhao, Magnetic skyrmion-based synaptic devices, Nanotechnology \textbf{28}, 08LT02 (2017).

\bibitem{Lisai_NANO2017} S. Li, W. Kang, Y. Huang, X. Zhang, Y. Zhou, and W. Zhao, Magnetic skyrmion-based artificial neuron device, Nanotechnology \textbf{28}, 31LT01 (2017).

\bibitem{Prychynenko_PRAPPL2018} D. Prychynenko, M. Sitte, K. Litzius, B. Kr{\"u}ger, G. Bourianoff, M. Kl{\"a}ui, J. Sinova, and K. Everschor-Sitte, Magnetic skyrmion as a nonlinear resistive element: a potential building block for reservoir computing, Phys. Rev. Appl. \textbf{9}, 014034 (2018).


\bibitem{Guoqiang_NL2017} G. Yu, P. Upadhyaya, Q. Shao, H. Wu, G. Yin, X. Li, C. He, W. Jiang, X. Han, P. K. Amiri, and K. L. Wang, Room-temperature skyrmion shift device for memory application, Nano Lett. \textbf{17}, 261 (2017).

\bibitem{Zazvorka_NN2019} J. Z\'{a}zvorka, F. Jakobs, D. Heinze, N. Keil, S. Kromin, S. Jaiswal, K. Litzius, G. Jakob, P. Virnau, D. Pinna, K. Everschor-Sitte, L. R\'{o}zsa, A. Donges, U. Nowak, and M. Kl\"{a}ui, Thermal skyrmion diffusion used in a reshuffler device, Nat. Nanotech. \textbf{14}, 658 (2019).

\bibitem{Woo_NE2020} K. M. Song, J.-S. Jeong, B. Pan, X. Zhang, J. Xia, S. Cha, T.-E. Park, K. Kim, S. Finizio, J. Raabe, J. Chang, Y. Zhou, W. Zhao, W. Kang, H. Ju, and S. Woo, Skyrmion-based artificial synapses for neuromorphic computing. Nat. Electron. \textbf{3}, 148 (2020).


\bibitem{Ezawa_PRB2011} M. Ezawa, Compact merons and skyrmions in thin chiral magnetic films, Phys. Rev. B \textbf{83}, 100408 (2011).

\bibitem{Lin_PRB2015} S.-Z. Lin, A. Saxena, and C. D. Batista, Skyrmion fractionalization and merons in chiral magnets with easy-plane anisotropy, Phys. Rev. B \textbf{91}, 224407 (2015).

\bibitem{Kharkov_PRL2017} Y. A. Kharkov, O. P. Sushkov, and M. Mostovoy, Bound states of skyrmions and merons near the Lifshitz point, Phys. Rev. Lett. \textbf{119}, 207201 (2017).

\bibitem{Leonov_PRB2017} A. O. Leonov and I. K\'{e}zsm\'{a}rki, Asymmetric isolated skyrmions in polar magnets with easy-plane anisotropy, Phys. Rev. B \textbf{96}, 014423 (2017).

\bibitem{Chmiel_NM2018} F. P. Chmiel, N. Waterfield Price, R. D. Johnson, A. D. Lamirand, J. Schad, G. van der Laan, D. T. Harris, J. Irwin, M. S. Rzchowski, C. B. Eom, and P. G. Radaelli, Observation of magnetic vortex pairs at room temperature in a planar $\alpha-\text{Fe}_2 \text{O}_3$/Co heterostructure, Nat. Mater. \textbf{17}, 581 (2018).

\bibitem{Kolesnikov_SR2018} A. G. Kolesnikov, V. S. Plotnikov, E. V. Pustovalov, A. S. Samardak, L. A. Chebotkevich, A. V. Ognev, and O. A. Tretiakov, Composite topological structure of domain walls in synthetic antiferromagnets, Sci. Rep. \textbf{8}, 15794 (2018).

\bibitem{Fernandes_SSC2019} R. L. Fernandes, R. J. C. Lopes, and A. R. Pereira, Skyrmions and merons in two-dimensional antiferromagnetic systems, Solid State Commun. \textbf{290}, 55 (2019).

\bibitem{Gobel_PRB2019}  B. G\"{o}bel, A. Mook, J. Henk, I. Mertig, and O. A. Tretiakov, Magnetic bimerons as skyrmion analogues in in-plane magnets, Phys. Rev. B \textbf{99}, 060407(R) (2019).

\bibitem{Kim_PRB2019} S. K. Kim, Dynamics of bimeron skyrmions in easy-plane magnets induced by a spin supercurrent, Phys. Rev. B \textbf{99}, 224406 (2019).

\bibitem{Moon_PRApplied2019} K.-W. Moon, J. Yoon, C. Kim, and C. Hwang, Existence of in-plane magnetic skyrmion and its motion under current flow, Phys. Rev. Applied \textbf{12}, 064054 (2019).

\bibitem{Murooka_SR2020} R. Murooka, A. O. Leonov, K. Inoue, and J.-I. Ohe, Current-induced shuttlecock-like movement of non-axisymmetric chiral skyrmions, Sci. Rep. \textbf{10}, 396 (2020).

\bibitem{Shen_PRL2020} L. Shen, J. Xia, X. Zhang, M. Ezawa, O. A. Tretiakov, X. Liu, G. Zhao, and Y. Zhou, Current-Induced Dynamics and Chaos of Antiferromagnetic Bimerons, Phys. Rev. Lett. \textbf{124}, 037202 (2020).

\bibitem{Lu_2020} X. Lu, R. Fei, and L. Yang, Meron-Like Topological Spin Defects in Monolayer CrCl$_{3}$, arXiv:2002.05208 (2020).

\bibitem{ES_PRB2020} R. Zarzuela, V. K. Bharadwaj, K. W. Kim, J. Sinova, and K. Everschor-Sitte, Stability and dynamics of in-plane skyrmions in collinear ferromagnets, Phys. Rev. B \textbf{101}, 054405 (2020).

\bibitem{DeAlfaro_PLB1976} V. De Alfaro, S. Fubini, and G. Furlan, A new classical solution of the Yang-Mills field equations, Phys. Lett. B \textbf{65}, 163 (1976).

\bibitem{Ezawa_2010} Z. F. Ezawa and G. Tsitsishvili, Skyrmion and bimeron excitations in bilayer quantum Hall systems, Physica E \textbf{42}, 1069 (2010).

\bibitem{Ezawa_2011} Z. F. Ezawa and G. Tsitsishvili, Skyrmion and bimeron excitations in imbalanced bilayer quantum Hall systems, AIP Conf. Proc. \textbf{1399}, 605 (2011).

\bibitem{Yu_Nature2018} X. Z. Yu, W. Koshibae, Y. Tokunaga, K. Shibata, Y. Taguchi, N. Nagaosa, and Y. Tokura, Transformation between meron and skyrmion topological spin textures in a chiral magnet, Nature \textbf{564}, 95 (2018).


\bibitem{Okubo_PRL2012} T. Okubo, S. Chung, and H. Kawamura, Multiple-q states and the skyrmion lattice of the triangular-lattice Heisenberg antiferromagnet under magnetic fields, Phys. Rev. Lett. \textbf{108}, 017206 (2012).

\bibitem{Leonov_NCOMMS2015} A. O. Leonov and M. Mostovoy, Multiply periodic states and isolated skyrmions in an anisotropic frustrated magnet, Nat. Commun. \textbf{6}, 8275 (2015).

\bibitem{Batista_Review2016} C. D. Batista, S.-Z. Lin, S. Hayami, and Y. Kamiya, Frustration and chiral orderings in correlated electron systems, Rep. Prog. Phys. \textbf{79}, 084504 (2016).

\bibitem{Lin_PRB2016A} S.-Z. Lin and S. Hayami, Ginzburg-Landau theory for skyrmions in inversion-symmetric magnets with competing interactions, Phys. Rev. B \textbf{93}, 064430 (2016).

\bibitem{Hayami_PRB2016A} S. Hayami, S.-Z. Lin, and C. D. Batista, Bubble and skyrmion crystals in frustrated magnets with easy-axis anisotropy, Phys. Rev. B \textbf{93}, 184413 (2016).

\bibitem{Rozsa_PRL2016} L. R{\'o}zsa, A. De{\'a}k, E. Simon, R. Yanes, L. Udvardi, L. Szunyogh, and U. Nowak, Skyrmions with attractive interactions in an ultrathin magnetic film, Phys. Rev. Lett. \textbf{117}, 157205 (2016).

\bibitem{Leonov_NCOMMS2017} A. O. Leonov and M. Mostovoy, Edge states and skyrmion dynamics in nanostripes of frustrated magnets, Nat. Commun. \textbf{8}, 14394 (2017).

\bibitem{Xichao_NCOMMS2017} X. Zhang, J. Xia, Y. Zhou, X. Liu, H. Zhang, and M. Ezawa, Skyrmion dynamics in a frustrated ferromagnetic film and current-induced helicity locking-unlocking transition, Nat. Commun. \textbf{8}, 1717 (2017).

\bibitem{Yuan_PRB2017} H. Y. Yuan, O. Gomonay, and M. Kl{\"a}ui, Skyrmions and multisublattice helical states in a frustrated chiral magnet, Phys. Rev. B \textbf{96}, 134415 (2017).

\bibitem{Hou_AM2017} Z. Hou, W. Ren, B. Ding, G. Xu, Y. Wang, B. Yang, Q. Zhang, Y. Zhang, E. Liu, F. Xu, W. Wang, G. Wu, X. Zhang, B. Shen, and Z. Zhang, Observation of various and spontaneous magnetic skyrmionic bubbles at room temperature in a frustrated kagome magnet with uniaxial magnetic anisotropy, Adv. Mater. \textbf{29}, 1701144 (2017).

\bibitem{Hu_SR2017} Y. Hu, X. Chi, X. Li, Y. Liu, and A. Du, Creation and annihilation of skyrmions in the frustrated magnets with competing exchange interactions, Sci. Rep. \textbf{7}, 16079 (2017).

\bibitem{Malottki_SR2017} S. von Malottki, B. Dupe, P. F. Bessarab, A. Delin, and S. Heinze, Enhanced skyrmion stability due to exchange frustration, Sci. Rep. \textbf{7}, 12299 (2017).

\bibitem{Liang_NJP2018} J. J. Liang, J. H. Yu, J. Chen, M. H. Qin, M. Zeng, X. B. Lu, X. S. Gao, and J. Liu, Magnetic field gradient driven dynamics of isolated skyrmions and antiskyrmions in frustrated magnets, New J Phys. \textbf{20}, 053037 (2018).

\bibitem{Diep_Entropy2019} T. H. Diep, Phase transition in frustrated magnetic thin film-physics at phase boundaries, Entropy \textbf{21}, 175 (2019).

\bibitem{Kurumaji_SCIENCE2019} T. Kurumaji, T. Nakajima, M. Hirschberger, A. Kikkawa, Y. Yamasaki, H. Sagayama, H. Nakao, Y. Taguchi, T.-h. Arima, and Y. Tokura, Skyrmion lattice with a giant topological Hall effect in a frustrated triangular-lattice magnet, Science \textbf{365}, 914 (2019).

\bibitem{Desplat_PRB2019} L. Desplat, J. V. Kim, and R. L. Stamps, Paths to annihilation of first- and second-order (anti)skyrmions via (anti)meron nucleation on the frustrated square lattice, Phys. Rev. B \textbf{99}, 174409 (2019).

\bibitem{Xia_PRApplied2019} J. Xia, X. Zhang, M. Ezawa, Z. Hou, W. Wang, X. Liu, and Y. Zhou, Current-driven dynamics of frustrated skyrmions in a synthetic aantiferromagnetic bilayer, Phys. Rev. Applied \textbf{11}, 044046 (2019).

\bibitem{Zarzuela_PRB2019} R. Zarzuela, H. Ochoa, and Y. Tserkovnyak, Hydrodynamics of three-dimensional skyrmions in frustrated magnets, Phys. Rev. B \textbf{100}, 054426 (2019).

\bibitem{Lohani_PRX2019} V. Lohani, C. Hickey, J. Masell, and A. Rosch, Quantum skyrmions in frustrated ferromagnets, Phys. Rev. X \textbf{9}, 041063 (2019).

\bibitem{Diep_2020} I. F. Sharafullin and H. T. Diep, Skyrmion Crystals and Phase Transitions in Magneto-Ferroelectric Superlattices: Dzyaloshinskii-Moriya Interaction in a Frustrated \textit{J}$_{1}$-\textit{J}$_{2}$ Model, Symmetry \textbf{12}, 26 (2020).


\bibitem{OOMMF} M. J. Donahue and D. G. Porter, Interagency Report NO. NISTIR 6376, National Institute of Standards and Technology, Gaithersburg, MD (1999) [http://math.nist.gov/oommf/].

\bibitem{Ado_PRB2017} I. A. Ado, O. A. Tretiakov, and M. Titov, Microscopic theory of spin-orbit torques in two dimensions, Phys. Rev. B \textbf{95}, 094401 (2017).


\bibitem{SM} See Supplemental Material at [\href{http://link.aps.org/supplemental/10.1103/PhysRevB.101.144435}{http://link.aps.org/supplemental-/10.1103/PhysRevB.101.144435}] for more information about the parameter dependency diagrams and dynamics simulation results.


\bibitem{Koshibae_NCOMMS2016} W. Koshibae and N. Nagaosa, Theory of antiskyrmions in magnets, Nat. Commun. \textbf{7}, 10542 (2016).

\bibitem{Ritzmann_NE2018} U. Ritzmann, S. von Malottki, J.-V. Kim, S. Heinze, J. Sinova, and B. Dup{\'e}, Trochoidal motion and pair generation in skyrmion and antiskyrmion dynamics under spin-orbit torques, Nat. Electron. \textbf{1}, 451 (2018).

\bibitem{Kovalev_Review2018} A. A. Kovalev and S. Sandhoefner, Skyrmions and antiskyrmions in quasi-two-dimensional magnets, Front. Phys. \textbf{6}, 98 (2018).

\bibitem{Potkina_2019} M. N. Potkina, I. S. Lobanov, O. A. Tretiakov, H. Jonsson, and V. M. Uzdin, Antiskyrmions in Ferromagnets and Antiferromagnets: Stability and Dynamics, arXiv:1906.06383 (2019).


\bibitem{Krane_1987} K. S. Krane, Introductory Nuclear Physics (Wiley, 1987).


\bibitem{Thiele_PRL1973} A. A. Thiele, Steady-state motion of magnetic domains, Phys. Rev. Lett. \textbf{30}, 230 (1973).

\bibitem{Tretiakov_2008A} O. A. Tretiakov, D. Clarke, G.-W. Chern, Ya. B. Bazaliy, and O. Tchernyshyov, Dynamics of domain walls in magnetic nanostrips, Phys. Rev. Lett. \textbf{100}, 127204 (2008).

\bibitem{Tretiakov_2008B} D. J. Clarke, O. A. Tretiakov, G.-W. Chern, Ya. B. Bazaliy, and O. Tchernyshyov, Dynamics of a vortex domain wall in a magnetic nanostrip: an application of the collective coordinate approach, Phys. Rev. B \textbf{78}, 134412 (2008).


\bibitem{Kaul_JMMM2004} E. E. Kaul, H. Rosner, N. Shannon, R. V. Shpanchenko, and C. Geibel, Evidence for a frustrated square lattice with ferromagnetic nearest-neighbor interaction in the new compound Pb$_2$VO(PO$_4$)$_2$, J. Magn. Magn. Mater. \textbf{272-276}, 922 (2004).

\end{thebibliography}
\end{document}